%% file: main.tex
\documentclass[letterpaper, 10 pt, conference]{ieeeconf}  

\IEEEoverridecommandlockouts                              
\usepackage[utf8]{inputenc}
\usepackage[T1]{fontenc}
\usepackage{booktabs}
\usepackage{graphicx}
\usepackage{placeins}
\usepackage{lipsum}
\usepackage{amsmath}
\usepackage{cite}
\usepackage{cleveref}
\usepackage{ragged2e}
\crefname{figure}{Fig.}{Figs.}
\Crefname{figure}{Fig.}{Figs.}
\Crefname{algorithm}{Algorithm}{Algorithms}
\usepackage{algorithm}
\usepackage{algpseudocode}
\usepackage{threeparttable}
\usepackage{multirow}
\usepackage{makecell}
\usepackage{placeins}

\title{\LARGE \bf
Functional Connectivity-Guided Band Selection for Motor Imagery Brain-Computer Interfaces
}

\author{Natália Araújo do Carmo$^{1}$ and Aarthy Nagarajan$^{2}$
\thanks{*This work was not supported by any organization}
\thanks{$^{1}$Natália Araújo do Carmo is with the College of Science,
        University of Notre Dame, Notre Dame, IN 46556, USA.
        Email: {\tt\small naraujod@nd.edu}}%
\thanks{$^{2}$Aarthy Nagarajan is with the College of Engineering and with the Lucy Family Institute for Data and Society,
       University of Notre Dame, Notre Dame, IN 46556, USA.
        Email: {\tt\small anagaraj@nd.edu} }   
}

\begin{document}

\maketitle
\thispagestyle{empty}
\pagestyle{empty}

\begin{abstract}

Reliable control in motor imagery brain-computer interfaces (MI-BCIs) requires the precise decoding of user-specific neural rhythms, which vary significantly across individuals. The Common Spatial Pattern (CSP) algorithm is a cornerstone of MI-BCI decoding, yet its performance depends strongly on the spectral range of the input EEG data. Although Filter Bank CSP (FBCSP) extends this as a data-driven decoding framework, its frequency sub-bands are predefined rather than selected using subject-specific physiological criteria. This paper presents a proof-of-concept study of static functional connectivity (FC)-guided band selection for MI-BCI, demonstrated using a conventional FBCSP-based pipeline. The proposed method identifies the most discriminative spectral bands by calculating phase-based connectivity across four sensorimotor channels using wPLI, PLV, and PLI. Nine bands in a 4-40 Hz filter bank are ranked by the effect size of their hemispheric coupling differences and pruned to the top $K$ bands for feature extraction and classification via FBCSP and a Support Vector Regressor. This framework was tested for $K$ values ranging from 1 to 8 across the BCI Competition IV-2a ($n=9$) and OpenBMI ($n=54$) datasets. Performance was benchmarked against standard nine-band FBCSP and random ablation to determine the minimum number of bands ($K^*$) required to maintain accuracy within a $\pm 2\%$ baseline equivalence zone. Results show FC-guided selection can outperform random ablation and achieve near-baseline performance while reducing required CSP fits by 22.2\% to 77.8\%. PLV enables the most aggressive dimensionality reduction by prioritizing the $\mu$ and low-$\beta$ ranges, while wPLI demonstrates superior inter-session robustness by mitigating volume conduction. These findings establish FC-guided selection as a principled and interpretable alternative to heuristic filter bank designs.

\end{abstract}

\section{INTRODUCTION}

Brain-computer interfaces (BCIs) are systems that acquire signals from neural activity and translate them into desired commands without using the usual output pathways of the nervous system \cite{wolpaw_braincomputer_2002}. This technology has a wide range of potential applications, from communication channels for patients with locked-in syndrome to advanced rehabilitation tools \cite{nicolas-alonso_brain_2012}. The basic structure of a BCI can be described as a series of steps: signal acquisition, feature extraction, feature translation, and the execution of commands by an output device, all of which are controlled by an operating protocol \cite{mcfarland_eeg-based_2017,wolpaw_braincomputer_2002}. Among the various available neuroimaging methods, electroencephalography (EEG) is the most commonly used for signal acquisition due to its non-invasive nature, portability, ease of use, and relatively inexpensive cost \cite{shih_brain-computer_2012}. EEG provides a direct measurement of neural activity with high temporal precision, but its spatial resolution is limited to synchronous populations of neurons whose electrical fields disperse through the scalp before detection by electrodes \cite{cohen_where_2017}. In addition, EEG is highly susceptible to biological artifacts such as electromyograms (EMGs) and electrooculograms (EOGs), which must be identified and removed prior to feature extraction \cite{mcfarland_brain-computer_2005}.

A highly effective paradigm for BCI control is motor imagery (MI), which corresponds to the mental process of imagining a movement without physically executing it. MI is particularly useful because it activates many of the same areas and pathways as physical movement execution \cite{pelgrims_contribution_2010}. Specifically, MI induces event-related desynchronizations (ERDs), which are reductions in the amplitude of $\mu$ (8-12 Hz) and $\beta$ (\textasciitilde{}13-30 Hz) rhythms in the sensorimotor cortex \cite{neuper_event-related_2001,barone_understanding_2021}. Successful extraction and decoding of these ERDs enables the creation of closed-loop MI-BCI systems \cite{shanechi_brainmachine_2019}. In these systems, user-specific algorithms translate MI into output commands while providing sensory feedback, thus stimulating bidirectional neural adaptation and activity-dependent plasticity that can be leveraged for control and rehabilitation \cite{muller-gerking_designing_1999,pfurtscheller_mu_2006,ang_eeg-based_2017}.

From a computational perspective, the success of an MI-BCI is highly dependent on its feature extraction and classification algorithms \cite{lotte_review_2018}. The Common Spatial Pattern (CSP) algorithm is a well-established and effective technique for this purpose \cite{muller-gerking_designing_1999,tangermann_review_2012, miladinovic_performance_2020}. CSP optimizes spatial filters to maximize the variance of EEG signals for one class of MI while minimizing it for another, making it highly effective in isolating ERD patterns. However, the efficacy of standard CSP is highly dependent on the spectral range of the input EEG data \cite{ang_filter_2012}. Because the reactive frequency ranges of $\mu$ and $\beta$ rhythms vary between individuals, applying a broad, standardized bandpass filter often degrades CSP performance.

To address CSP's spectral selectivity, the Filter Bank Common Spatial Pattern (FBCSP) algorithm was developed \cite{ang_filter_2008}. FBCSP bandpass-filters EEG into multiple frequency bands, extracts CSP features from each of them, and then selects the most discriminative band-CSP pairs across the spectrum. Despite its widespread adoption and success as a data-driven framework, standard FBCSP has an important limitation: the sub-bands in its filter-bank are predefined rather than selected using subject-specific physiological criteria \cite{darvishi_eeg-driven_2025,lee_eeg_2019}. This presents an opportunity to streamline the feature extraction pipeline. By incorporating subject-specific neurophysiological characteristics earlier in the process, it may be possible to target the most discriminative bands before spatial filtering even occurs.

In this paper, we present a proof-of-concept study of static phase-based functional connectivity (FC)-guided band selection for MI-BCI, demonstrated using a conventional FBCSP pipeline. Phase-based FC captures the temporal synchronization between spatially distributed neural populations \cite{marzetti_brain_2019,daly_brain_2012}. Metrics such as the weighted Phase Lag Index (wPLI), Phase Locking Value (PLV), and Phase Lag Index (PLI) quantify the degree to which different regions of the sensorimotor network communicate during motor imagery tasks \cite{vinck_improved_2011, lachaux_measuring_1999, bruna_phase_2018,stam_phase_2007}. By evaluating phase synchrony, FC adds a network-level dimension of neurophysiological information that can facilitate the selection of the most discriminative spectral bands in a filter bank.

Although existing literature has explored functional connectivity in MI-BCIs, FC is most commonly utilized as an additional feature family \cite{wang_diverse_2020,li_common_2019,darvishi_eeg-driven_2025, siviero_functional_2023,ai_feature_2019}. In these paradigms, FC metrics are typically concatenated with CSP spatial features to form an expanded, multi-modal feature vector prior to classification. While this approach successfully integrates complementary neural information, it inherently increases the dimensionality of the overall feature space, which may introduce additional computational demands and increase the risk of overfitting the classifier.

To leverage functional connectivity without inflating the feature space, we propose using FC as a neurophysiologically informed, subject-specific band selection criterion. This approach identifies spectral bands exhibiting MI-related network connectivity to rank and prune the filter bank prior to FBCSP spatial filtering. This targeted selection paradigm allows the FBCSP algorithm to process the most highly synchronized frequencies, thus reducing the required spatial filter optimizations. 

The FC-based band selection framework was evaluated in the BCI Competition IV-2a ($n=9$) and OpenBMI ($n=54$) datasets. The contributions of this work are as follows:

\begin{itemize}
    \item We introduce a novel, neurophysiologically grounded, and subject-specific band selection method that utilizes phase-based functional connectivity to rank and prune frequency bands prior to FBCSP spatial filtering.
    
    \item We demonstrate that FC-guided band selection can reduce the required spatial filter optimizations by 22.2 to 77.8\% while maintaining classification accuracy within a $\pm 2\%$ equivalence zone of a standard nine-band FBCSP baseline.

    \item We statistically verify that FC-guided pruning significantly outperforms random ablation in several evaluated configurations, most notably within PLV-based pipelines.

    \item We provide a systematic comparison of phase-based metrics, suggesting that PLV maximizes dimensionality reduction by consistently selecting canonical $\mu$ and low-$\beta$ rhythms, while wPLI offers superior inter-session robustness by mitigating volume conduction.
\end{itemize}


\section{METHODOLOGY}

\subsection{Datasets}

Two publicly available datasets were used to measure the performance of the proposed method.

\subsubsection{BCI Competition IV Dataset 2a}

For initial evaluations, we used Dataset 2a \textit{Continuous Multi-class Motor Imagery} from BCI Competition IV \cite{tangermann_review_2012}. EEG signals were recorded from nine healthy subjects using 22 Ag/AgCl electrodes. Each participant performed four-class motor imagery (left hand, right hand, feet, and tongue) in two separate sessions on different days. Signals were captured at a sampling rate of 250 Hz and processed using a 0.5-100 Hz bandpass filter. Three EOG channels were also sampled at 250 Hz.

This study utilized only the first session, as it contained class labels for all trials. Data loading was limited to left- and right-hand trials as the proposed method is focused on binary motor imagery classification. Epochs were extracted over a window from 0.5 to 2.5 seconds after the onset of the visual cue as presented in the original FBCSP implementation \cite{ang_filter_2008}. Signals from the EOG channels were removed to reduce noise. This dataset will be referred to as BCIC IV-2a.

\subsubsection{Korea University OpenBMI}

In this study, only the MI paradigm from the Korea University OpenBMI dataset \cite{lee_eeg_2019} was used. EEG signals were recorded from 54 healthy subjects using 62 Ag/AgCl electrodes placed according to the International 10-20 system. Each participant performed binary-class MI (left and right hand) in two sessions. Signals were captured at a sampling rate of 1,000 Hz. Four EMG channels were also sampled at 1,000 Hz.

To reduce computational demand, the EEG signals were downsampled to 250 Hz. Data loading was limited to 21 channels within the sensorimotor cortex (Fz, FC5, FC3, FC1, FC2, FC4, FC6, C5, C3, C1, Cz, C2, C4, C6, CP5, CP3, CP1, CPz, CP2, CP4, CP6). Epochs were extracted over a window from 1.0 to 3.5 seconds after visual cue onset to avoid initial visual evoked potentials from stimulus presentation and to exclude late-trial cognitive anticipation, thereby isolating the most stable period of MI. The two sessions were evaluated both independently and in combination. This dataset will be referred to as OpenBMI.

\subsection{Proposed Method}

To establish functional connectivity as a viable mechanism for subject-specific frequency band selection, this proof-of-concept study integrates an FC-guided selection framework into a standard FBCSP pipeline, as illustrated in \Cref{alg:proposed_method}. The FBCSP component was implemented using the reference code from \cite{ang_filter_2008}.

The method was evaluated using a 10-fold cross-validation framework. To prevent information leakage, the FC-based frequency band ranking and selection were performed strictly within the training set of each fold. In the first stage, data were partitioned into training and testing sets. The training data were then decomposed into nine frequency bands ranging from 4 to 40 Hz at 4 Hz intervals.

\input{alg_pipeline}

\subsubsection{Channel Selection}

Four EEG channels were selected to compute functional connectivity. In both datasets, three primary motor cortex channels were utilized: C3 (left hemisphere), Cz (midline), and C4 (right hemisphere). To capture frontal midline activity, FCz was used for BCIC IV-2a; however, due to its absence in the OpenBMI dataset, Fz was selected. 

\subsubsection{Frequency Band Ranking and Selection}

\begin{figure}[htbp]
        \centering
        \includegraphics[width=\columnwidth]{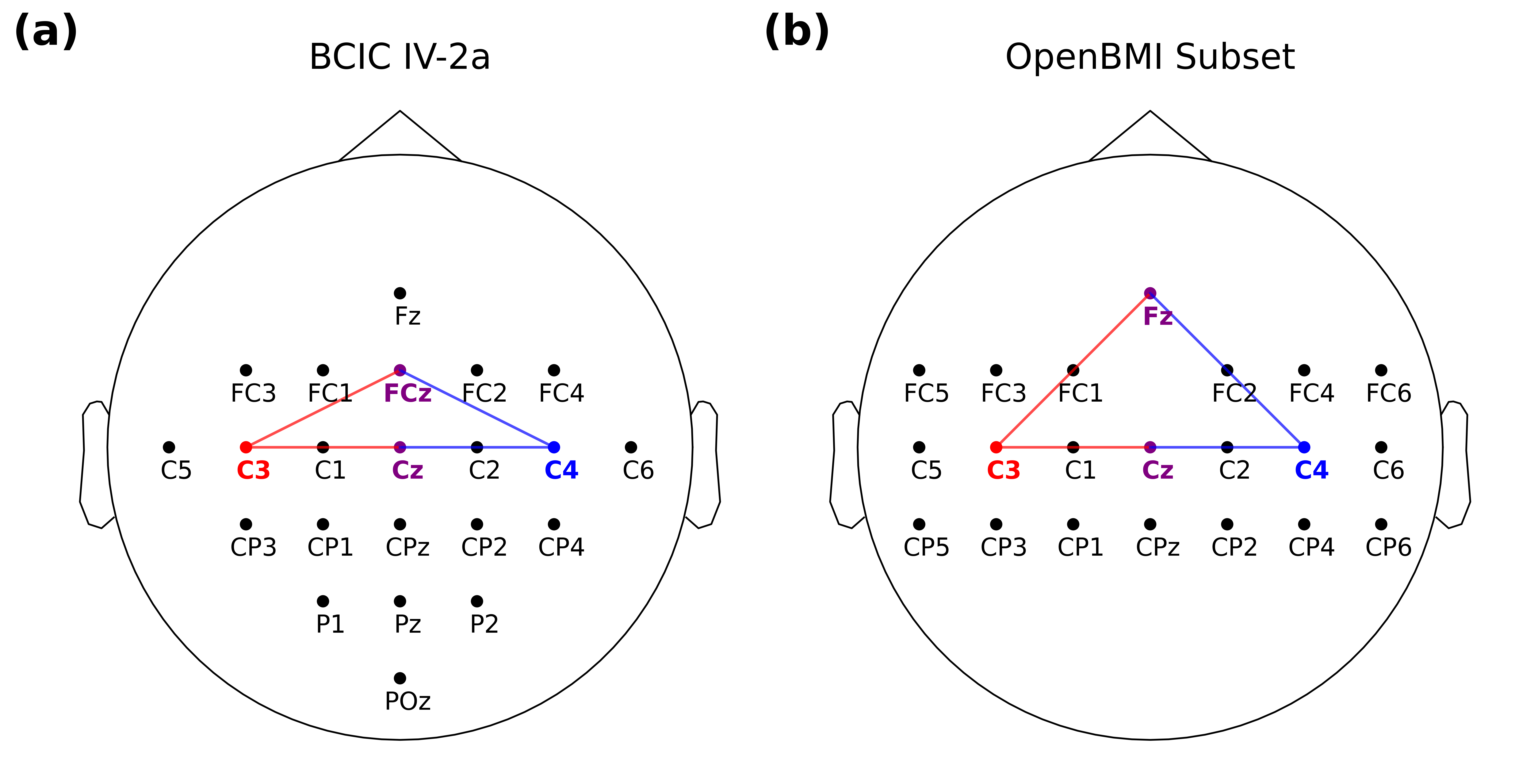}
        \caption{Spatial distribution of EEG channels included in the experimental pipeline for the \textbf{(a)} BCIC IV-2a and \textbf{(b)} OpenBMI datasets. The specific channels and edges (pairwise connections) utilized for FC calculations are highlighted in red for the left hemisphere, and blue for the right hemisphere. The evaluated edges were C3-Cz and C3-FCz/Fz for the left hemisphere, and C4-Cz and C4-FCz/Fz for the right. Because FCz was unavailable in the OpenBMI dataset, Fz was used in its place.}
        \label{fig:fc_layout}
    \end{figure}

For each frequency band, functional connectivity scores were calculated using the \texttt{spectral\_connectivity\_time} function from the MNE-Connectivity package \cite{larson_mne-python_2024,gramfort_meg_2013,li_mne-connectivity_2026}. As illustrated in \cref{fig:fc_layout}, connectivity was evaluated in pairwise fashion: C3-Cz, C3-FCz/Fz, C4-Cz, and C4-FCz/Fz.

The primary metric used for functional connectivity calculations was weighted Phase Lag Index (wPLI), with Phase Locking Value (PLV) and Phase Lag Index (PLI) included as exploratory comparisons. wPLI was chosen as the primary metric because it best balances robustness to volume conduction with sensitivity to non-zero-lag interactions \cite{vinck_improved_2011}. PLV was included as a liberal benchmark; although highly susceptible to zero-lag volume conduction, it remains sensitive to MI-related synchrony and has been widely used in existing MI-BCI research \cite{wang_phase_2006,wang_diverse_2020,darvishi_eeg-driven_2025,gonuguntla_phase_2013}. Conversely, PLI was utilized as a strictly conservative baseline that strongly suppresses volume conduction, verifying the robustness of connectivity patterns at the cost of reduced sensitivity \cite{leeuwis_functional_2021,stam_phase_2007}.

Following the computation of functional connectivity scores, the training data were grouped by MI class (left- vs. right-hand). For every trial within each class, the connectivity scores were then averaged across the left hemisphere channel pairs and right hemisphere channel pairs, yielding a mean connectivity score for each hemisphere. Subsequently, for every trial within each MI class, a coupling difference value ($\Delta C$) was defined as:
$$ \Delta C = C_{\text{L}} - C_{\text{R}} $$
where $C_{\text{L}}$ and $C_{\text{R}}$ represent the mean connectivity of the left and right hemispheres, respectively. Under a simple lateralization assumption based on the contralateral nature of motor control, left- and right-hand MI might be expected to yield negative and positive $\Delta C$ values, respectively. However, because this strict sign convention is not always guaranteed in phase-based FC, our approach relied on the magnitude of class separation rather than on the specific sign of $\Delta C$. For each band, the feature evaluated for class discrimination was the per-trial hemispheric coupling difference $\Delta C$. To quantify how strongly the distributions of $\Delta C$ differ between left- and right-hand MI, the magnitude of the effect size was calculated using a modified Cohen's $d$:
$$d = \frac{|M_\text{L} - M_\text{R}|}{S_\text{pooled} + \epsilon}$$
where $M_\text{L}$ and $M_\text{R}$ represent the mean $\Delta C$ for left- and right-hand MI, respectively, and $S_\text{pooled}$ denotes the pooled standard deviation computed as $\sqrt{(s_{\text{L}}^2 + s_{\text{R}}^2) / 2}$. A small constant, $\epsilon = 10^{-12}$ was added to the denominator to ensure numerical stability by preventing division by zero. By calculating the absolute value in the numerator, the final band score $d$ captures the magnitude of the effect. Thus, larger $d$ scores imply bands that are more discriminative between MI classes.
Within each fold, the nine frequency bands were evaluated and ranked in descending order based on their calculated $d$ score. Finally, the top $K$ most discriminative bands were selected to be passed on to the FBCSP algorithm, where $K$ serves as the experimental hyperparameter governing the number of selected bands. 

\subsubsection{Feature Extraction and Classification}

The FBCSP algorithm \cite{ang_filter_2008} was then applied to the selected subset of $K$ frequency bands to extract discriminative, subject-specific spatial features. The extracted features were passed to a Support Vector Regressor (SVR) implemented in scikit-learn \cite{scikit-learn}, utilizing a Radial Basis Function (RBF) kernel. The kernel coefficient was set to $\gamma= \text{'auto'}$, while all other hyperparameters were kept at their default values. Finally, a mean testing accuracy was computed across the 10 folds to evaluate overall performance.

\subsection{Experimental Design}

To evaluate the proposed method, independent experimental runs were conducted across all subjects using each of the three FC metrics (wPLI, PLV, and PLI). For OpenBMI, the two recording sessions had separate experimental runs.
For each FC metric, the hyperparameter $K$ (the number of selected frequency bands) was evaluated iteratively from $K=1$ to $K=8$. Furthermore, each configuration was tested both with and without the Mutual Information-based Best Individual Feature (MIBIF) algorithm \cite{ang_filter_2008} selecting for four features. While the FC-guided band selection acts as a macro-level filter to identify neurophysiologically relevant frequency bands, FBCSP still extracts multiple spatial features per band. MIBIF acts as a secondary, micro-level noise reduction layer to select the four most discriminative CSP features. The pipeline was tested with and without MIBIF to demonstrate whether the proposed static FC-guided band selection is powerful enough to stand on its own or if it works best as a two-stage feature selection pipeline with MIBIF.

\subsubsection{Baselines and Ablation Studies}

To demonstrate that FC-guided band selection can lead to comparable classification accuracy with a reduced feature space and to verify its physiological relevance, the proposed method was benchmarked against two comparators:
\begin{itemize}
    \item \textbf{Standard FBCSP Baseline:} A standard nine-band FBCSP pipeline utilizing all frequency bands in the 4-40 Hz range was evaluated both with and without MIBIF to establish the standard performance baseline.
    \item \textbf{Random Selection Ablation:} To ensure that the performance of the proposed method was explicitly due to FC guidance rather than simply a reduced parameter space, a random ablation study was conducted with and without MIBIF. For each value of $K$ and each fold, bands were selected at random over 20 iterations. The classification accuracies from these 20 random draws were averaged to yield a single representative accuracy per fold. Then, the classification accuracies were averaged across the 10 folds to determine the subject-level performance.
\end{itemize}

\subsection{Statistical Analysis}

\subsubsection{Comparison to Baseline}

The primary goal of the analysis was to identify the minimum number of frequency bands $K^{*}$ needed to reach an equivalent mean classification accuracy to the nine-band baseline. To achieve this, we calculated the accuracy delta for each $K$, which was defined as the difference between the mean accuracy at that $K$ and the mean baseline accuracy such that $\Delta=\overline{Acc}_{K} - \overline{Acc}_{baseline}$. Thus, a negative delta means below-baseline performance. We established an equivalence zone of $\pm 2\%$ relative to the mean baseline accuracy and determined that an equivalent performance was reached if the entire 95\% confidence interval (CI) of the mean fell within the $\pm 2\%$ equivalence zone. The 95\% CIs were calculated from the subject-level accuracy deltas using a nonparametric percentile bootstrap method with 9,999 resamples.

\subsubsection{Comparison to Random Ablation}

To determine whether the proposed FC-guided selection significantly outperformed random chance, classification performance for each value of $K$ was compared against the random ablation using a one-sided Wilcoxon signed-rank test. To control for the family-wise error rate across multiple comparisons, all $p$-values were adjusted using the Holm-Bonferroni correction.

\subsubsection{Quantification of Computational Efficiency}

To quantify the efficiency gained through FC-guided band selection, computational savings were evaluated based on the reduction in spatial filter optimizations. The baseline FBCSP pipeline processes nine frequency bands, while our proposed FC-guided method restricts feature extraction only to the top $K^*$ bands. Consequently, this approach reduces the training compute load by requiring fewer CSP fits and generating a smaller intermediate feature space. The number of CSP filter pairs was fixed at $m=2$ for all experiments, so the intermediate feature space was reduced from the 36-feature baseline down to $4K^*$. In configurations where MIBIF was utilized, this space was further reduced to 4 features.

\subsubsection{Band Selection Stability}

To evaluate the stability of the FC-guided band selection, the mean selection frequency of each frequency band was analyzed. The selection ratio of each of the nine bands across the 10 cross-validation folds was computed per subject and subsequently averaged over $K=1$ through $8$. Because the selection ratios are bounded continuous variables, non-parametric statistical methods were employed. A Friedman test was first conducted to determine if there were significant differences in selection frequency across the nine bands. If statistical significance was found ($p<0.05$), post-hoc pairwise comparisons were performed using Wilcoxon signed-rank tests. To control for Type I error rates across multiple comparisons, all post-hoc $p$-values were adjusted using the Benjamini-Hochberg False Discovery Rate (FDR) correction.

\subsection{Neurophysiological Interpretability}

To evaluate the neurophysiological relevance of the frequency bands selected by the FC-guided pipeline, we visualized the spatial activation patterns of the best- and worst-performing subjects in the OpenBMI dataset. We focused on the two most robust metrics identified in our analysis: wPLI and PLV. Subjects were evaluated strictly within a single configuration (Session 2, $K=7$ bands, without MIBIF) to ensure an independent evaluation without the confounding effects of secondary feature selection. For these subjects, the "Top" and "Worst" bands were defined as the most and least frequently selected bands, respectively, across the 10 cross-validation folds.

We extracted the spatial patterns from the CSP filters using the Haufe transform, defined as:
$$A = \Sigma_x W \Sigma_{\hat{s}}^{-1}$$
where $\Sigma_x$ is the covariance matrix of the filtered EEG data, $W$ is the unmixing matrix of CSP filters, and $\Sigma_{\hat{s}}$ is the covariance matrix of the latent sources \cite{haufe_interpretation_2014}. This transforms the backward decoding weights into physiologically interpretable source activations. Computations were performed using the \texttt{mne.decoding.CSP} module in MNE-Python \cite{larson_mne-python_2024}, which natively extracts these forward-model patterns.

For each cross-validation fold, the data were filtered to the target band and CSP was fitted strictly on the training epochs. Because the mathematical sign of CSP eigenvectors is arbitrary, directly averaging patterns across folds causes destructive interference. We resolved this using a sign-alignment protocol: the spatial pattern from the first fold served as a template, and for subsequent folds, the pattern's polarity was inverted if its Pearson correlation coefficient with the template was negative. Furthermore, because CSP components do not guarantee a fixed class order, we explicitly mapped components to specific motor tasks by projecting the training data through the fitted filters and computing the class-conditional variances. The component maximizing the variance ratio between left-hand and right-hand MI was assigned to left-hand MI, while the component minimizing this ratio was assigned to right-hand MI. Finally, the aggregated spatial patterns were plotted with dynamically computed, strictly symmetric colormap limits centered at zero.

\section{RESULTS}

\subsection{Baseline Classification Performance} 

The nine-band baseline classification accuracies for each dataset are presented in \Cref{tab:baseline_accuracy}. The baseline accuracy was 84.10\% $\pm$ 16.59\% for BCIC IV-2a with MIBIF, 85.41\% $\pm$ 14.59\% for BCIC IV-2a without MIBIF, 68.60\% $\pm$ 17.61\% for OpenBMI with MIBIF, and 67.61\% $\pm$ 17.68\% for OpenBMI without MIBIF. The large standard deviations and the difference in means across the two datasets reveal the high inter-subject variability in MI-BCI performance. In addition, the application of MIBIF resulted in marginal differences in average accuracy across both datasets. Given the large standard deviations, MIBIF does not appear to provide a substantial performance gain, suggesting it serves primarily as a secondary dimensionality reduction tool rather than a primary driver of baseline accuracy in these FBCSP pipelines.

\begin{table}[htbp]
    \centering
    \caption{Nine-band Baseline Accuracy (\%) (mean $\pm$ SD).}
    \label{tab:baseline_accuracy}
    \input{tables/baseline_accuracy}
\end{table}

\subsection{Comparison with Baseline and Ablation}

The following analyses have two primary objectives: first, to determine the minimum number of frequency bands ($K^{*}$) required to retain near-baseline classification performance, and second, to establish whether the proposed FC-guided selection significantly outperforms random band ablation.

\begin{figure}[htbp]
        \centering
        \includegraphics[width=\columnwidth]{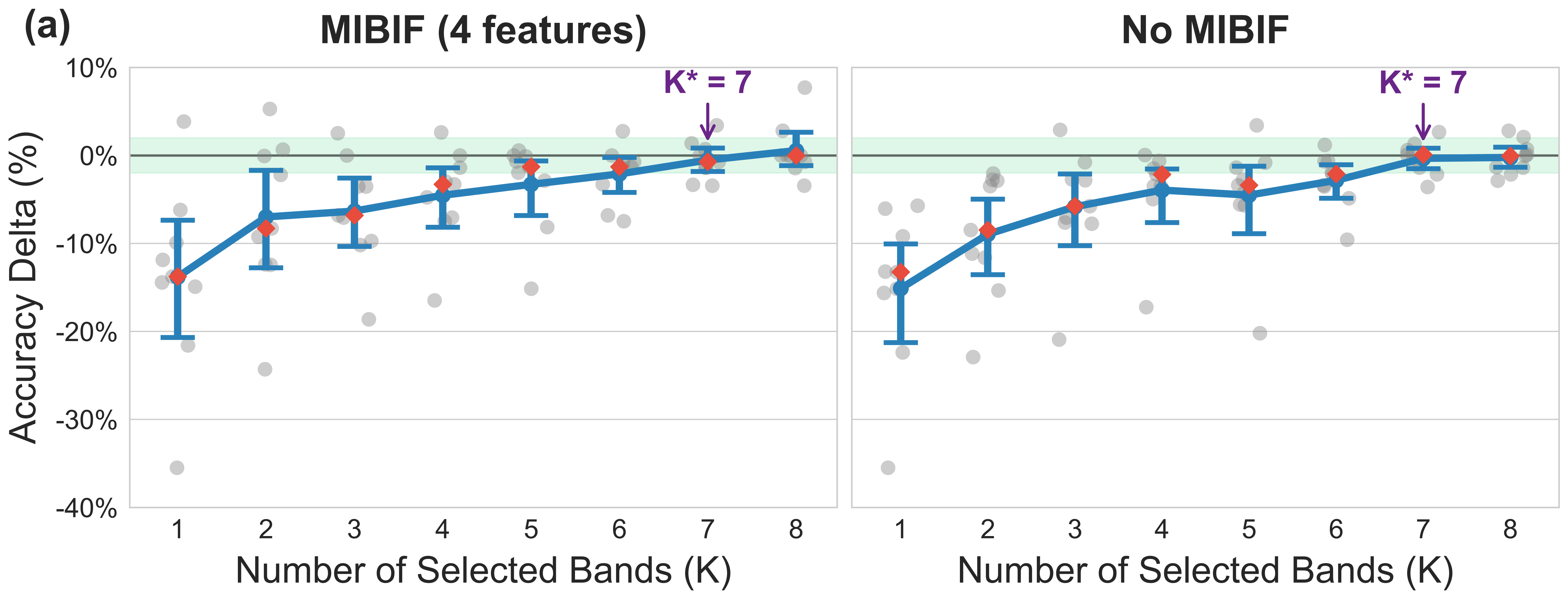}
        \vspace{0.2cm}
        \includegraphics[width=\columnwidth]{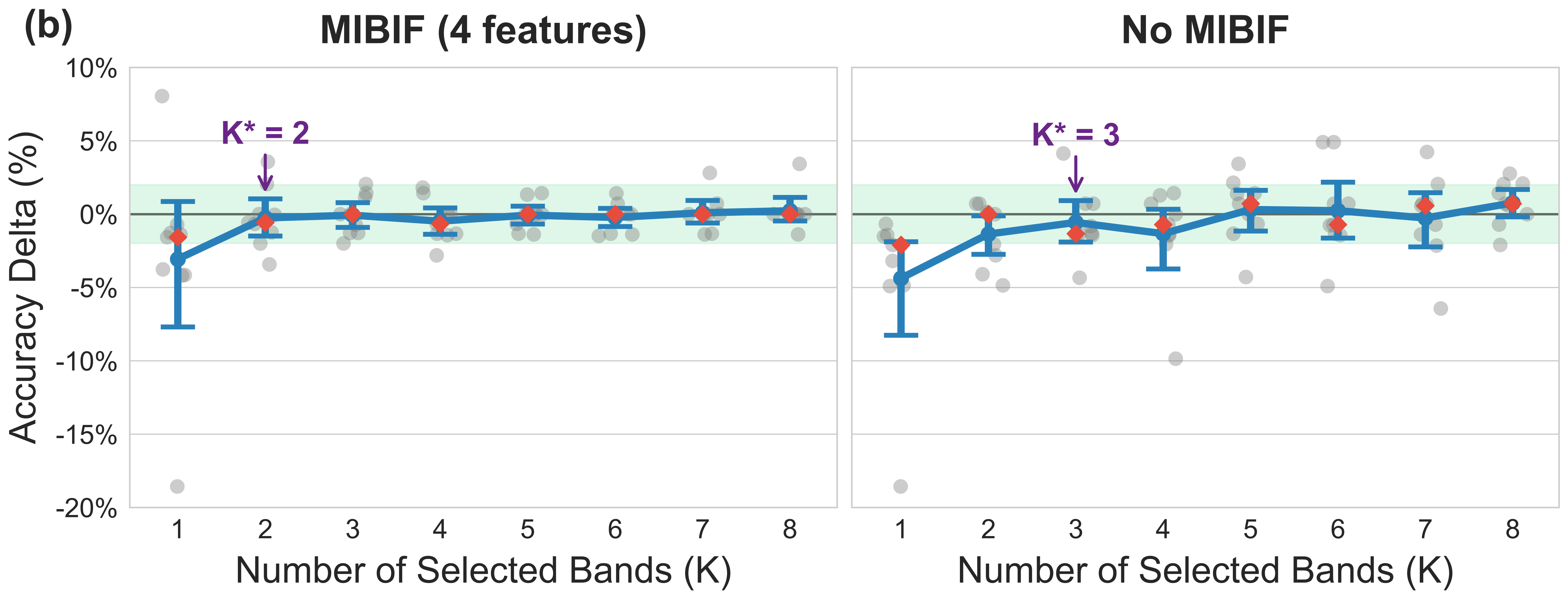}
        \vspace{0.2cm}
        \includegraphics[width=\columnwidth]{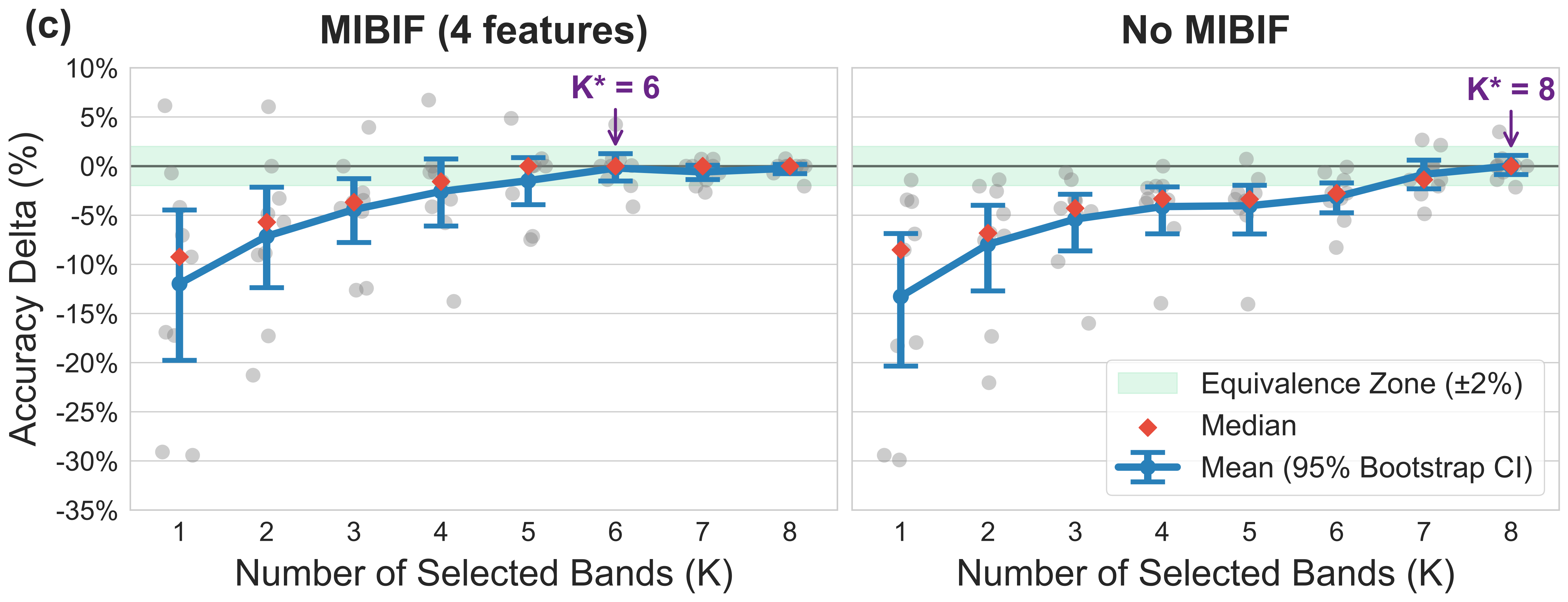}      
        \caption{Accuracy delta as a function of the number of selected frequency bands ($K$) across three FC metrics: \textbf{(a)} wPLI, \textbf{(b)} PLV, and \textbf{(c)} PLI with BCIC IV-2a data ($n=9$ subjects). Each panel compares the classification performance using pipelines with and without MIBIF selection of four features. For a given $K$, gray points represent subject-specific accuracies, while red markers and blue points with error bars denote the mean and median (with 95\% bootstrap CIs) accuracies, respectively. The green shaded region indicates the $\pm 2\%$ baseline equivalence zone. Purple arrows highlight $K^*$, defined as the smallest $K$ where the CI of the mean delta is fully within the equivalence zone. The lowest values of $K^*$ are obtained using PLV.}
        \label{fig:topk_bcic}
    \end{figure}

As demonstrated in \Cref{tab:mink_bcic}, the PLV-guided method reached baseline equivalence retaining only 2 to 3 frequency bands ($K^*=2$ with MIBIF, $K^*=3$ without) in the BCIC IV-2a dataset. In contrast, PLI and wPLI required retaining nearly the entire filter bank (6 to 8 bands) to achieve near-baseline performance. Furthermore, \Cref{fig:topk_bcic} illustrates that the mean classification accuracy with PLV never decreased by more than 5\% from the baseline, even at $K=1$, while PLI and wPLI exhibited a greater reduction in accuracy at low $K$. PLV thus achieved the baseline equivalence plateau much faster than the other metrics. This rapid stabilization highlights the parsimony and non-inferiority of the PLV-guided approach rather than superiority over the baseline, as it allows for aggressive dimensionality reduction while retaining near-baseline performance.

    \begin{table}[htbp]
        \centering
        \caption{Minimum Bands ($K^*$) for Baseline Equivalence \protect\\ (BCIC IV-2a, $n=9$)}
        \label{tab:mink_bcic}
        \input{tables/mink_bcic_table}
        
        \vspace{2pt}
        {\justifying\footnotesize\noindent The $^*$ indicates configurations where the proposed method significantly outperformed the random ablation ($p_{\text{adj}}<0.05$), as detailed in \cref{tab:bcic_wilcoxon_ablation}. The $^\dagger$ indicates that MIBIF was used to select 4 features. Bolded $K^*$ value denotes the minimum value of $K^*$ achieved.\par}
    \end{table}

\begin{figure}[htbp]
        \centering
        \includegraphics[width=\columnwidth]{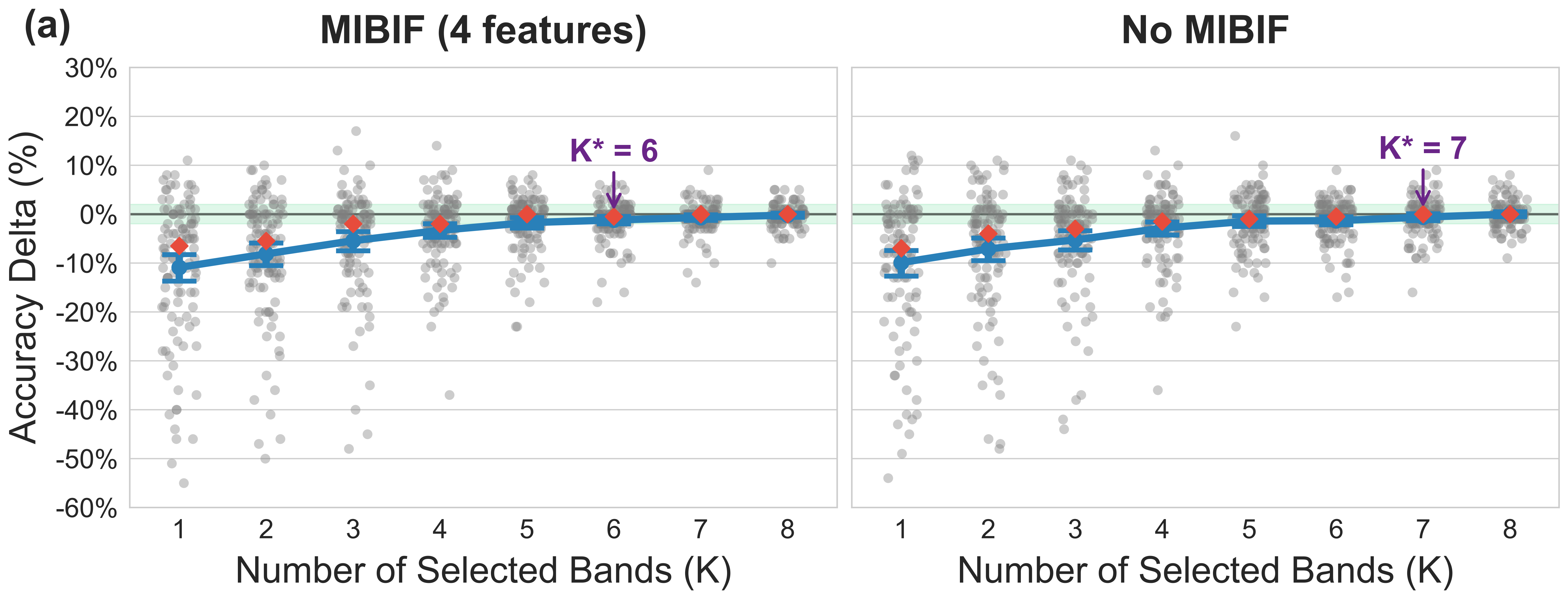}
        
        \vspace{0.2cm}
        
        \includegraphics[width=\columnwidth]{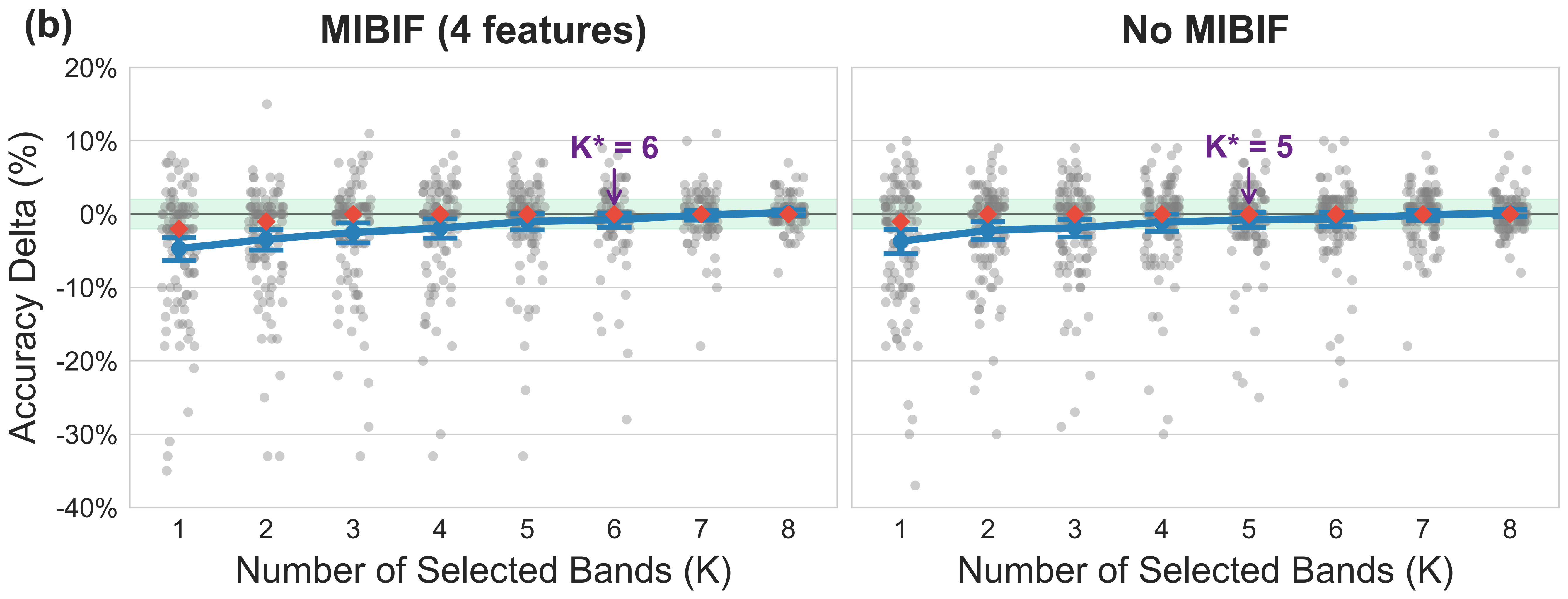}
        
        \vspace{0.2cm}
        
        \includegraphics[width=\columnwidth]{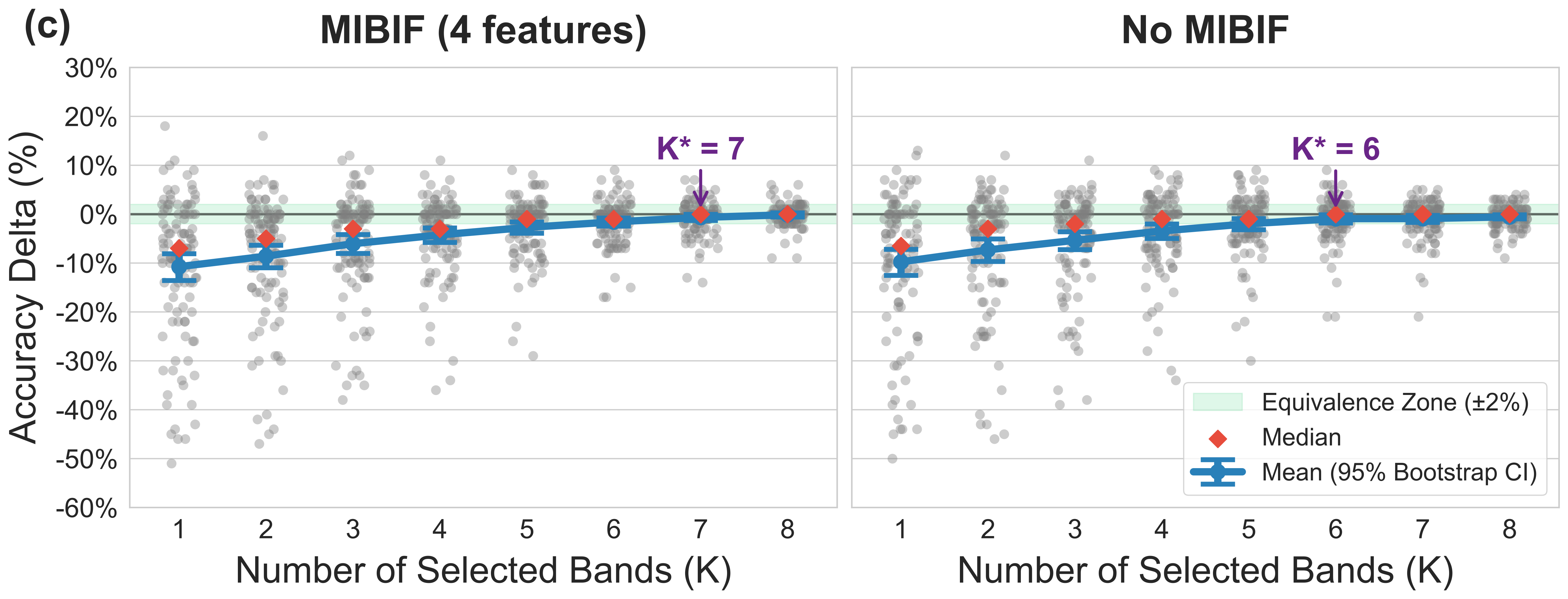}   
        
        \caption{Accuracy delta as a function of the number of selected frequency bands ($K$) across three FC metrics: \textbf{(a)} wPLI, \textbf{(b)} PLV, and \textbf{(c)} PLI with the combined sessions of OpenBMI data ($n=54$ subjects). Each panel compares the classification performance using pipelines with and without MIBIF selection of four features. For a given $K$, gray points represent subject-specific accuracies, while red markers and blue points with error bars denote the mean and median (with 95\% bootstrap CIs) accuracies, respectively. The green shaded region indicates the $\pm 2\%$ baseline equivalence zone. Purple arrows highlight $K^*$, defined as the smallest $K$ where the CI of the mean delta is fully within the equivalence zone. The lowest value of $K^*$ is obtained with PLV without MIBIF.}
        \label{fig:topk_openbmi}
    \end{figure}

This trend generalizes to the larger OpenBMI dataset, although less dramatically (\Cref{tab:mink_openbmi}). Notably, the minimum number of bands ($K^*$) required to reach baseline equivalence with the PLV metric is higher for OpenBMI compared to BCIC IV-2a (increasing from 2 to 3 bands up to 5 to 6 bands). While wPLI and PLI required a consistently high number of bands across both datasets (6 to 8), PLV's ability to aggressively prune spectral bands was attenuated in the larger cohort. This difference likely reflects the increased inter-subject variability and broader signal diversity inherent in a much larger cohort ($n=54$ versus $n=9$), requiring the retention of more spectral information to maintain near-baseline performance at the population level. Despite this attenuation, the PLV-guided method again reached baseline equivalence using the fewest bands ($K^*=6$ with MIBIF, $K^*=5$ without for the combined sessions), compared to the 6 to 7 bands required by wPLI and PLI. Similarly, the mean classification accuracy with PLV never decreased by more than 5\% from baseline, even at $K=1$, whereas PLI and wPLI exhibited more substantial reduction in accuracy at low $K$ (\Cref{fig:topk_openbmi}). However, when analyzing performance by individual session, wPLI demonstrated the highest inter-session stability, as $K^*$ either decreased or remained constant from session 1 to session 2 (\Cref{tab:mink_openbmi}).

    \begin{table}[htbp]
        \centering
        \caption{Minimum Bands ($K^*$) for Baseline Equivalence \protect\\ (OpenBMI, $n=54$)}
        \label{tab:mink_openbmi}
        \input{tables/mink_openbmi_table}
        
        \vspace{2pt}
        {\justifying\footnotesize\noindent The $^*$ and $^{**}$ indicate configurations where the proposed method significantly outperformed the random ablation, where $^*$: $p_{\text{adj}}<0.05$ and $^{**}$: $p_{\text{adj}}<0.001$, as detailed in \cref{tab:openbmi_wilcoxon_ablation}. The $^\dagger$ indicates that MIBIF was used to select 4 features. Bolded $K^*$ values denote the minimum value of $K^*$ achieved within each respective session.\par}
    \end{table}

On the BCIC IV-2a dataset, FC-guided selection strictly dominated random ablation only when it aggressively pruned the bands (\Cref{tab:bcic_wilcoxon_ablation}). The PLV method with MIBIF ($K^*=2$) achieved 83.84\% accuracy compared to a random accuracy of 73.21\% ($p_{\text{adj}}<0.05$), a performance advantage of 10.63\%. Without MIBIF ($K^*=3$), PLV had an accuracy of 84.84\% compared to a random accuracy of 76.45\% ($p_{\text{adj}}<0.05$), a performance advantage of 8.39\%. For wPLI and PLI, where $K^*$ remained high (6 to 8), performance did not significantly differ from random ablation. At high $K^*$ values, random selection is nearly mathematically identical to FC-guided selection, causing the performance of the proposed method and random ablation to naturally converge.

    \begin{table}[htbp]
        \centering
        \caption{Statistical Comparison of the Proposed Method vs. Random Ablation at $K^*$ (BCIC IV-2a, $n=9$)}
        \label{tab:bcic_wilcoxon_ablation}
        \input{tables/bcic_wilcoxon_ablation}
        
        \vspace{2pt}
        {\justifying\footnotesize\noindent Results are based on a one-sided Wilcoxon signed-rank test ($H_1: \text{Method} > \text{Random}$). The reported $p$-values were adjusted for multiple comparisons using the Holm-Bonferroni method. The $^*$ denotes statistical significance ($p_{\text{adj}}<0.05$). The $^\dagger$ indicates that MIBIF was used to select 4 features.\par}
    \end{table}

In the combined OpenBMI sessions, FC-guided selection significantly outperformed random selection for all methods except wPLI without MIBIF (\Cref{tab:openbmi_wilcoxon_ablation}). PLV demonstrated the highest margins of improvement over random ablation (1.91\%, $p_{\text{adj}}<0.01$ with MIBIF, and 2.70\%, $p_{\text{adj}}<0.01$ without). Although the accuracy advantages were narrower on this dataset, the highly significant $p$-values suggest that FC-guided selection consistently extracts more informative features compared to random chance across a large subject pool.

    \begin{table}[htbp]
        \centering
        \caption{Statistical Comparison of the Proposed Method vs. Random Ablation at $K^*$ (OpenBMI, $n=54$)}
        \label{tab:openbmi_wilcoxon_ablation}
        \input{tables/openbmi_wilcoxon_ablation}
        
        \vspace{2pt}
        {\justifying\footnotesize\noindent Results are based on a one-sided Wilcoxon signed-rank test ($H_1: \text{Method} > \text{Random}$). The reported $p$-values were adjusted for multiple comparisons using the Holm-Bonferroni method. The $^*$ and $^{**}$ denote statistical significance, where $^*$: $p_{\text{adj}}<0.05$ and $^{**}$: $p_{\text{adj}}<0.001$. The $^\dagger$ indicates that MIBIF was used to select 4 features.\par}
    \end{table}

Ultimately, the ability to retain a small number of informative frequency bands through FC-guided selection directly translates into substantial computational savings during model training and feature extraction. \Cref{tab:compute_savings} details the relative computational savings achieved by the proposed method with the configurations that demonstrated a statistically significant performance advantage over random band ablation. This ensures that the reported computational efficiency is driven by the extraction of informative functional connectivity rather than arbitrary dimensionality reduction.

In the BCIC IV-2a dataset, PLV was the only metric that significantly outperformed random ablation. By isolating the most discriminative information into just 2 or 3 bands, PLV greatly reduced the spatial filtering computations of FBCSP, achieving a 77.8\% reduction in required CSP fits when paired with MIBIF ($K^*=2$) and a 66.7\% reduction by itself ($K^*=3$). In the OpenBMI dataset, while multiple FC metrics yielded statistically significant classification accuracies at $K^*$, PLV enabled the most substantial compute savings. It provided a 44.4\% reduction in CSP training computations without MIBIF ($K^*=5$), surpassing the savings offered by wPLI and PLI. When paired with MIBIF, PLV provided an equal reduction in CSP fits as wPLI with MIBIF and PLI without MIBIF, achieving 33.3\% in compute savings ($K^*=6$). The lowest reduction in CSP training computations was 22.2\%, which was achieved by PLI with MIBIF ($K^*=7$). These results demonstrate that FC-guided band selection can reduce the computational costs of model training while preserving classification accuracy.

    \begin{table}[htbp]
        \centering
        \caption{Computational Savings for Configurations Outperforming Random Ablation}
        \label{tab:compute_savings}
        \input{tables/compute_savings}

        \vspace{2pt}
        {\justifying\footnotesize\noindent This table exclusively reports configurations demonstrating significant performance gains over random ablation (detailed in \Cref{tab:bcic_wilcoxon_ablation,tab:openbmi_wilcoxon_ablation}). Computational savings indicate the percentage reduction in CSP filter optimizations versus the nine-band baseline, defined explicitly as $100 \times (9-K^*)/9$. The $^\dagger$ indicates that MIBIF was used to select 4 features.\par}
    \end{table}

\subsection{Band Selection Stability}

To investigate whether the FC metrics consistently prioritize the same neurophysiological features, we analyzed band selection stability across cross-validation folds and subjects. This analysis specifically focused on session 2 of the OpenBMI dataset ($n=54$). Session 2 was chosen because it represents an independent evaluation phase recorded on a separate day, providing a more rigorous representation of potential real-world testing conditions. We selected the wPLI and PLV metrics for this analysis since wPLI demonstrated the highest inter-session stability and PLV consistently achieved the lowest required band threshold ($K^*$).

\begin{figure}[htbp]
        \centering
        \includegraphics[width=\columnwidth]{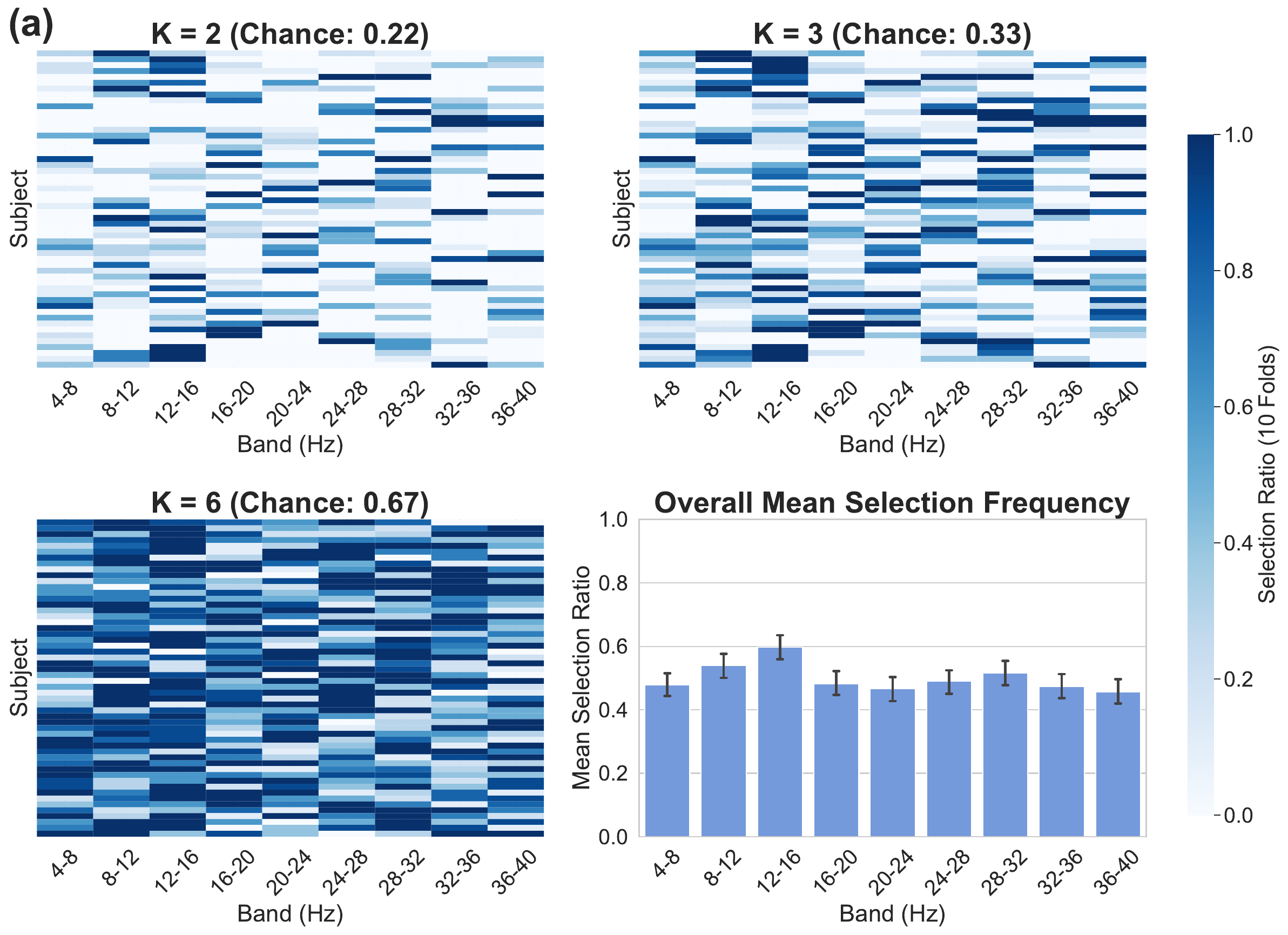}
        \vspace{0.15cm}
        \includegraphics[width=\columnwidth]{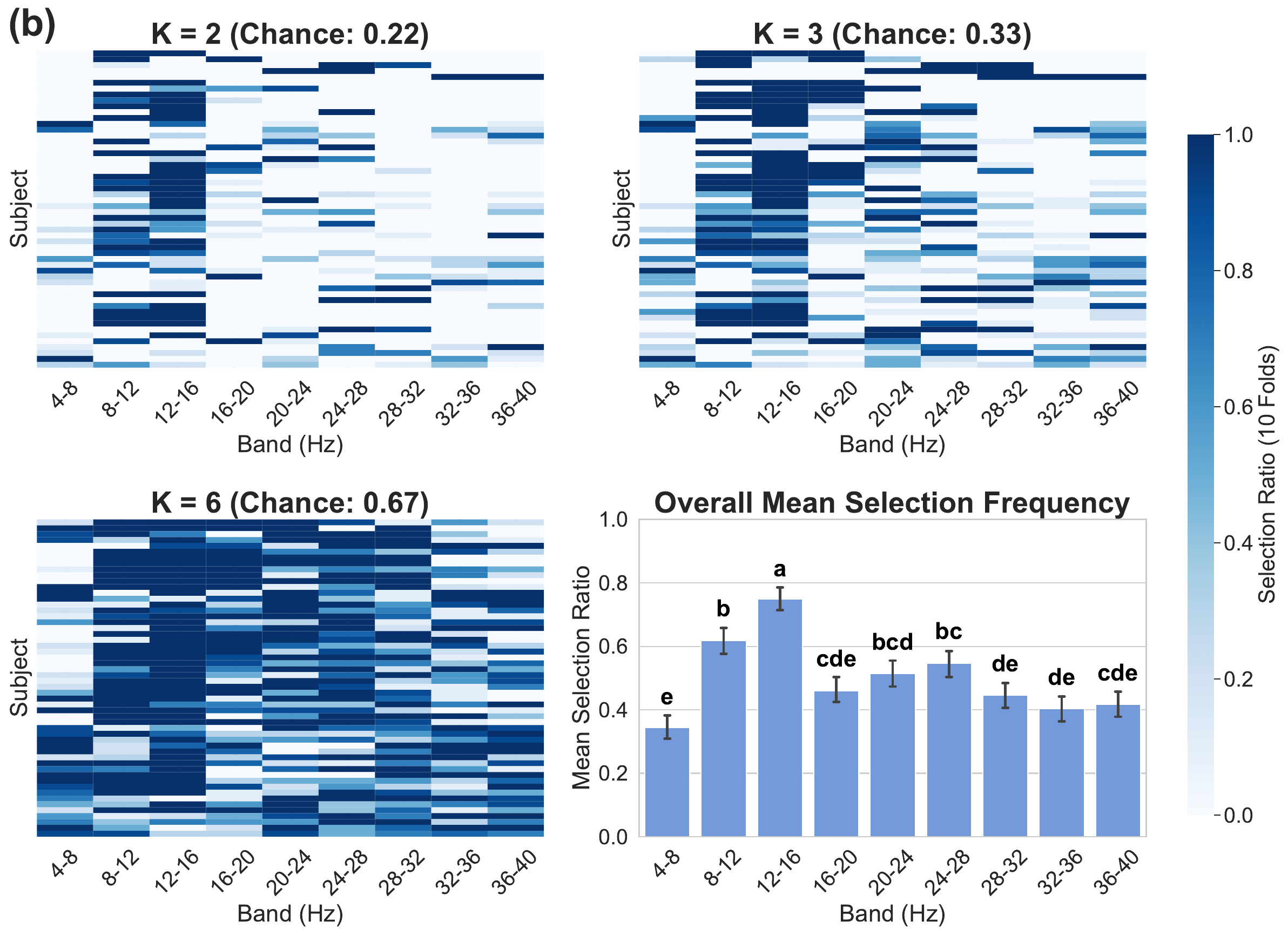}     
        \caption{Stability of frequency band selection guided by \textbf{(a)} wPLI and \textbf{(b)} PLV with session 2 of OpenBMI ($n=54$ subjects). wPLI was selected for highest inter-session stability and PLV was selected for lowest $K^*$. Session 2 was selected because it represents an independent evaluation phase. The first three panels illustrate subject-level band selection ratios across 10 cross-validation folds for $K=2, 3,$ and $6$. Subjects ($n=54$) on the y-axis are sorted in descending order based on their baseline classification accuracy in session 2. The bottom-right panel displays the overall mean selection frequency for each band, aggregated across $K=1$ through $8$. Error bars denote 95\% CIs. Pairwise statistical differences are denoted using a compact letter display above the error bars. Frequency bands that do not share a common letter are significantly different ($p_{\text{adj}}<0.05$), assessed via a Friedman test followed by post-hoc FDR-corrected Wilcoxon signed-rank tests.}
        \label{fig:band_selection_stability}
    \end{figure}

\Cref{fig:band_selection_stability} illustrates these selection patterns by comparing both the subject-level and population-level consistency of band selection with wPLI and PLV alongside their overall mean selection frequencies. The heatmaps highlight band selection ratios at $K=2, 3,$ and $6$ to provide a representative view of band prioritization across different levels of dimensionality reduction. Specifically, $K=2$ and $K=3$ demonstrate band selection stability where selections are furthest from random chance. The value $K=6$ then illustrates how those selection patterns evolve as the subsets approach the optimal $K^*$ thresholds.

Analysis of the overall selection frequencies revealed a distinct disparity in band prioritization between the two metrics. The Friedman test indicated no significant difference in the overall selection frequency of bands using wPLI-guided selection ($\chi^2(8)=10.06$, $p=0.261$). This lack of significance suggests a highly dispersed selection pattern across the 4-40 Hz frequency spectrum. In contrast, PLV exhibited highly significant differences in band preference ($\chi^2(8)=55.77$, $p<0.001$). Post-hoc FDR-corrected Wilcoxon signed-rank tests revealed that the 12-16 Hz band was selected significantly more often than all eight other frequency bands ($p_{\text{adj}}<0.05$ for all pairwise comparisons). Additionally, the 8-12 Hz band was selected significantly more often than the lower 4-8 Hz band and most of the higher frequency bands above 16 Hz. The complete list of post-hoc pairwise comparisons for PLV is detailed in Appendix \Cref{tab:plv_posthoc}. These results collectively indicate that PLV selection is heavily concentrated in the $\mu$ (8-12 Hz) and low-$\beta$ (12-16 Hz) ranges, whereas wPLI selection is much more dispersed. This concentrated selection suggests that the PLV metric tends to prioritize frequency bands traditionally associated with motor imagery \cite{pfurtscheller_motor_1997,neuper_event-related_2001,jeon_event-related_2011}. Furthermore, qualitative observation of the heatmaps in \Cref{fig:band_selection_stability} reinforces this population-level stability. For PLV, high selection ratios visually align vertically across the most frequently selected bands for the majority of subjects. The wPLI heatmaps instead show dark cells spread irregularly across the 4-40 Hz frequency spectrum. PLV also demonstrates higher intra-subject stability by repeatedly selecting the same specific bands across folds for individual subjects more consistently than wPLI.

\subsection{Neurophysiological Interpretability}

To validate that the FC-guided pipeline selects frequency bands with interpretable neurophysiological relevance, we visualized the spatial activation patterns for the best- and worst-performing subjects in the OpenBMI dataset. We continued to focus on the wPLI and PLV metrics. To strictly isolate the pipeline's raw feature-selection capabilities on an evaluation session, we restricted this visualization to session 2 without MIBIF. Under these specific conditions, both metrics had $K^*=7$, which was also selected for visualization. Because both FC metrics yielded two subjects with perfect (100.0\%) classification accuracy with the chosen configurations, Subject 36 was randomly selected to represent the top performers for both. Conversely, the worst-performing subjects achieved below-chance accuracies of 40.0\% for PLV and 47.0\% for wPLI.

\begin{figure}[htbp]
         \centering
         \includegraphics[width=\columnwidth]{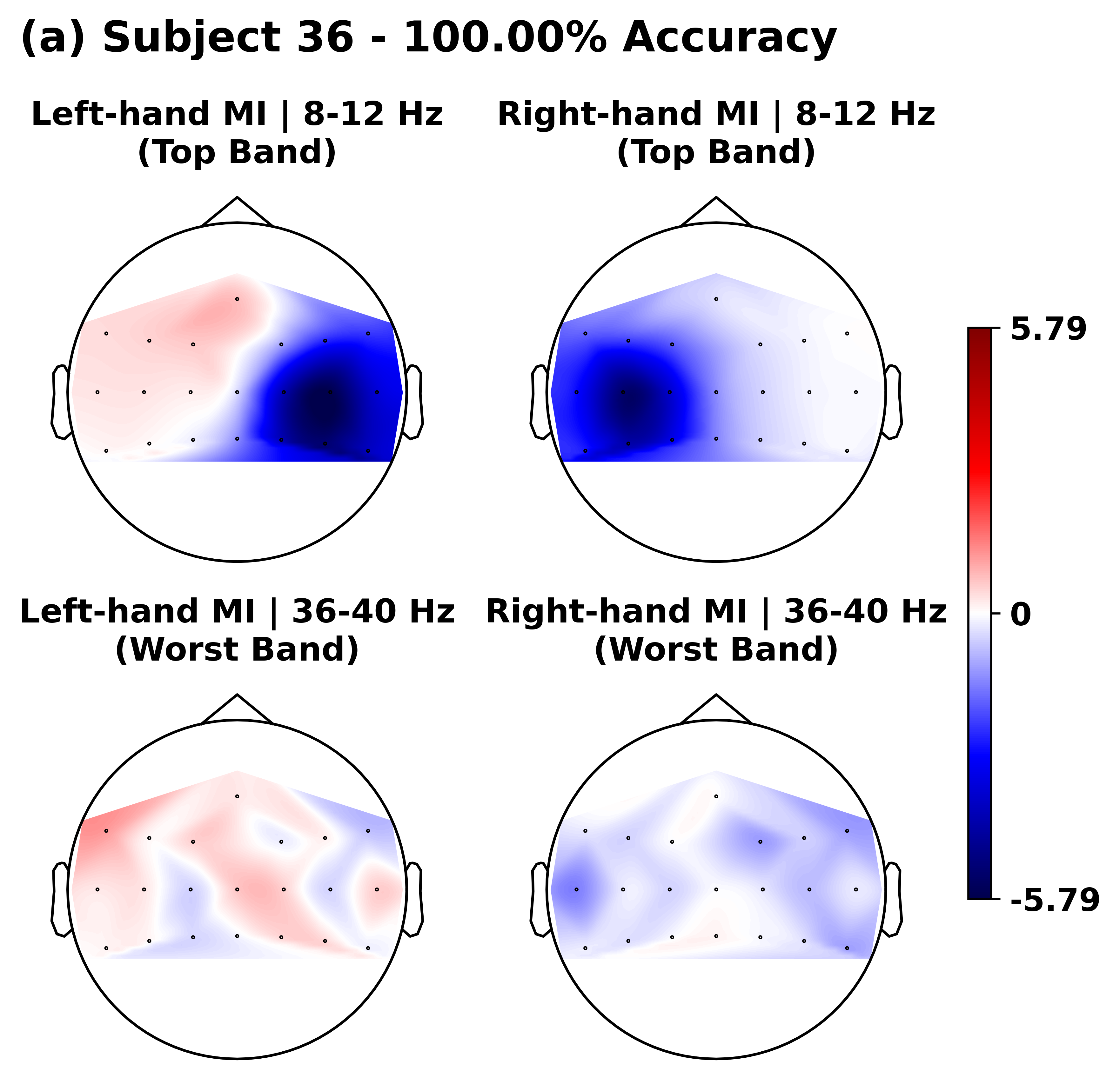}
         
         \vspace{0.25cm}
         
         \includegraphics[width=\columnwidth]{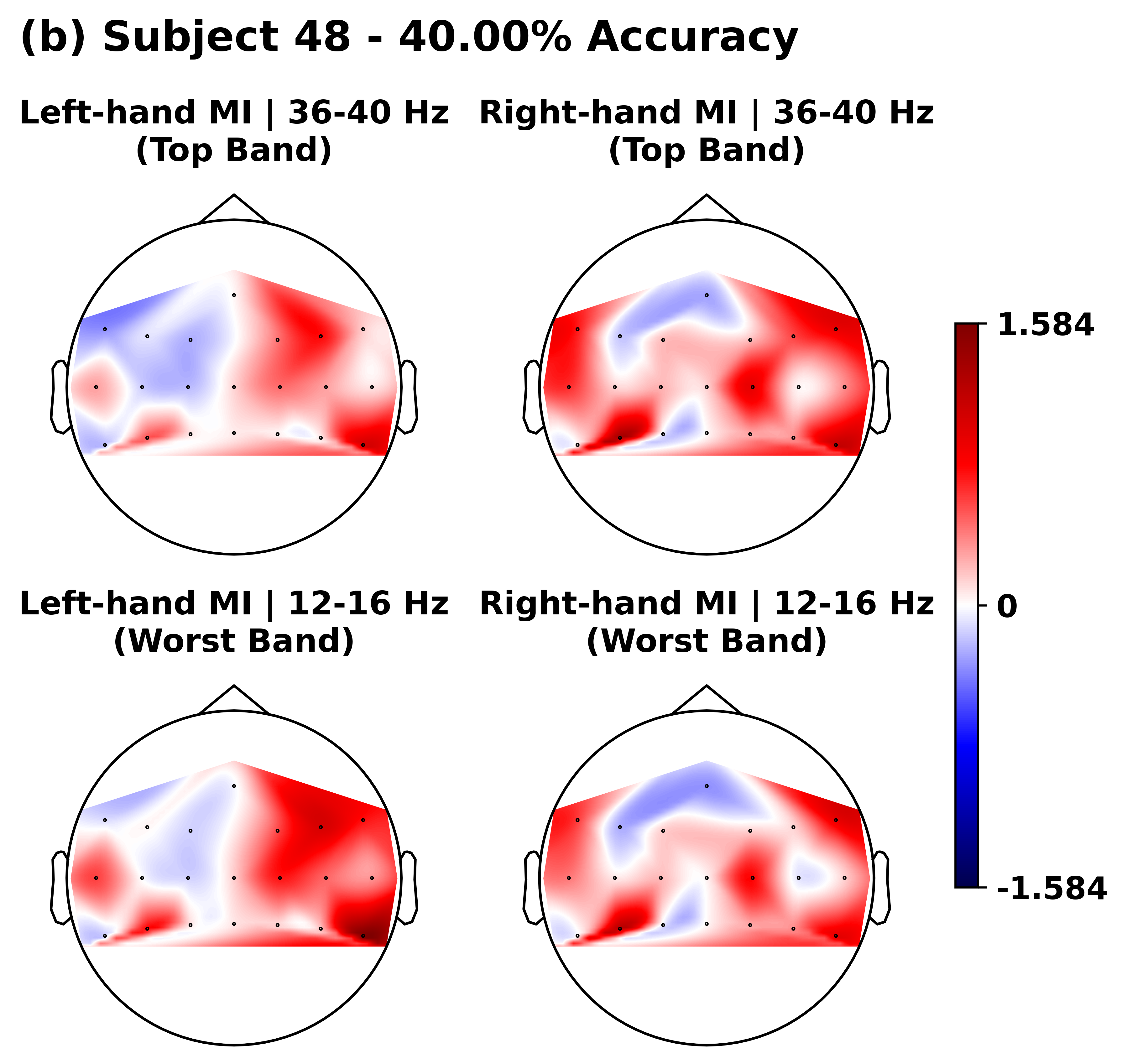}
         \caption{PLV topomaps for the OpenBMI dataset ($n=54$) during session 2. Panels depict the subjects with the \textbf{(a)} highest (Subject 36) and \textbf{(b)} lowest (Subject 48) mean classification accuracies, evaluated at $K^*=7$ with no MIBIF. Topomaps visualize CSP spatial patterns. For each specified frequency band, a CSP model was fit across all trials to distinguish left- and right-hand motor imagery. Color gradients correspond to the spatial weights of the extreme CSP components (representing the distinct classes) mapped onto 21 selected motor cortex electrode locations. Within each panel, the upper and lower topomaps display the most and least frequently selected frequency bands, respectively. Band selection frequencies were aggregated across the 10 cross-validation folds. The upper topomaps display the most frequently selected band for each subject: 8-12 Hz (selected 100.0\% of the time) for Subject 36, and 36-40 Hz (100.0\%) for Subject 48. The lower topomaps display the least frequently selected band for each subject: 36-40 Hz (0.0\%) for Subject 36, and 12-16 Hz (30.0\%) for Subject 48.}
         \label{fig:topomaps_plv}
     \end{figure}

PLV is sensitive to all phase-locked activity, including zero-lag connectivity, which makes it well-suited for capturing highly synchronized, localized cortical dipoles. This is evident in the spatial patterns of the best-performing PLV subject (\Cref{fig:topomaps_plv}(a)). For this subject, PLV-guided selection identified the 8-12 Hz band ($\mu$) as a discriminative feature in 100.0\% of the cross-validation folds. The resulting topographies revealed dense focal activation over the contralateral sensorimotor cortices for left- and right-hand MI, centered around the C4 and C3 electrodes, respectively. Furthermore, the pipeline successfully ignored uninformative features: the 36–40 Hz band ($\gamma$-range) was not selected in any of the CV folds and exhibited only faint, dispersed topographic patterns.

This stands in contrast with the worst-performing PLV subject (\Cref{fig:topomaps_plv}(b)), whose lack of clear sensorimotor lateralization is visually apparent. Here, the pipeline defaulted to a high-frequency range, selecting the 36–40 Hz band ($\gamma$-range) 100.0\% of the time. The resulting spatial patterns are bilateral, scattered, and lack any C3/C4 lateralization. This indicates that in the absence of a strong task-modulated $\mu$-rhythm, the classifier was forced to rely on non-informative high-frequency noise, resulting in chance-level performance. These findings indicate that PLV's primary strength lies in its ability to identify and leverage highly lateralized sensorimotor $\mu$-rhythms to achieve superior performance and interpretability.

\begin{figure}[htbp]
         \centering
         \includegraphics[width=\columnwidth]{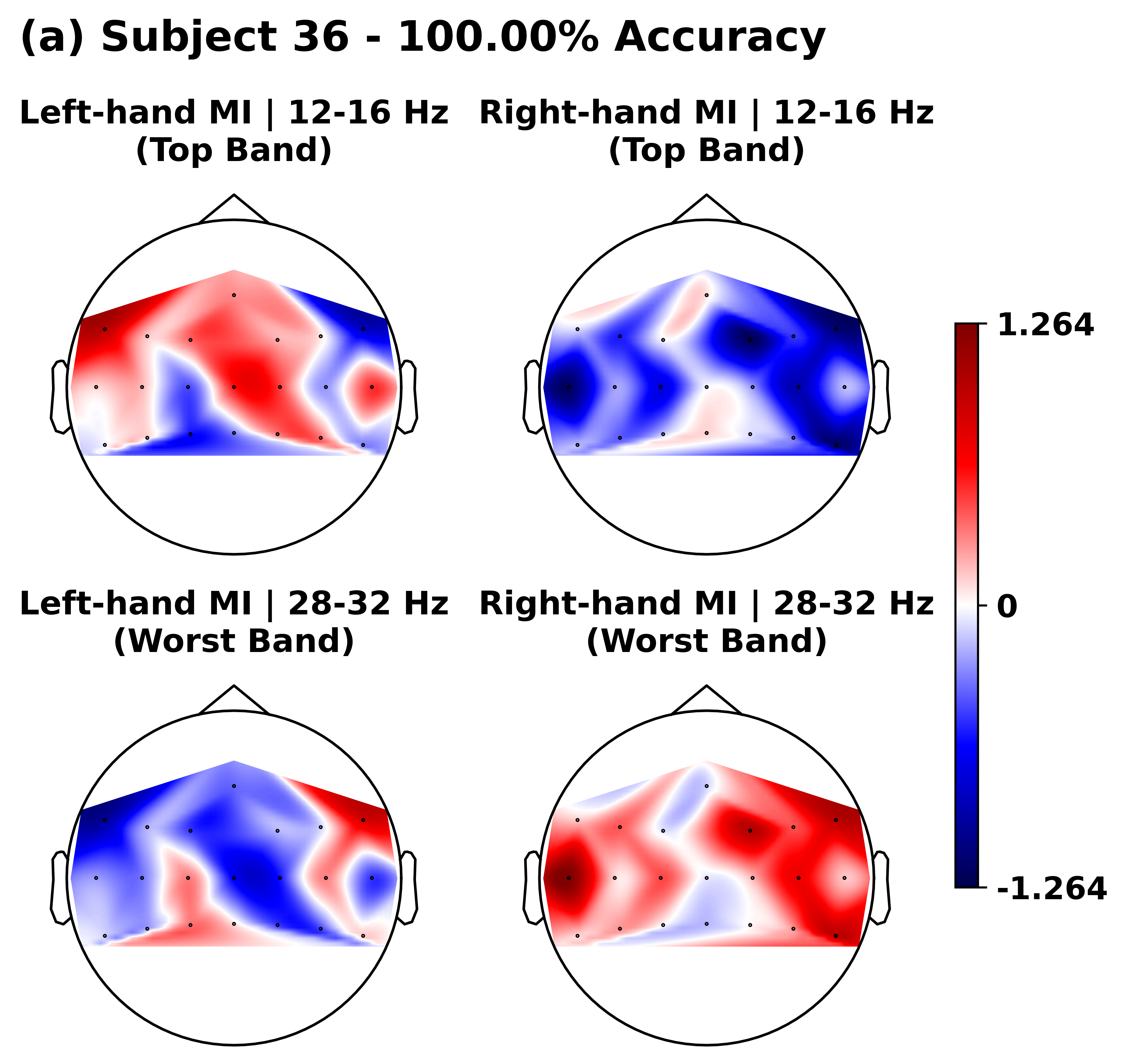}
         
         \vspace{0.25cm}
         
         \includegraphics[width=\columnwidth]{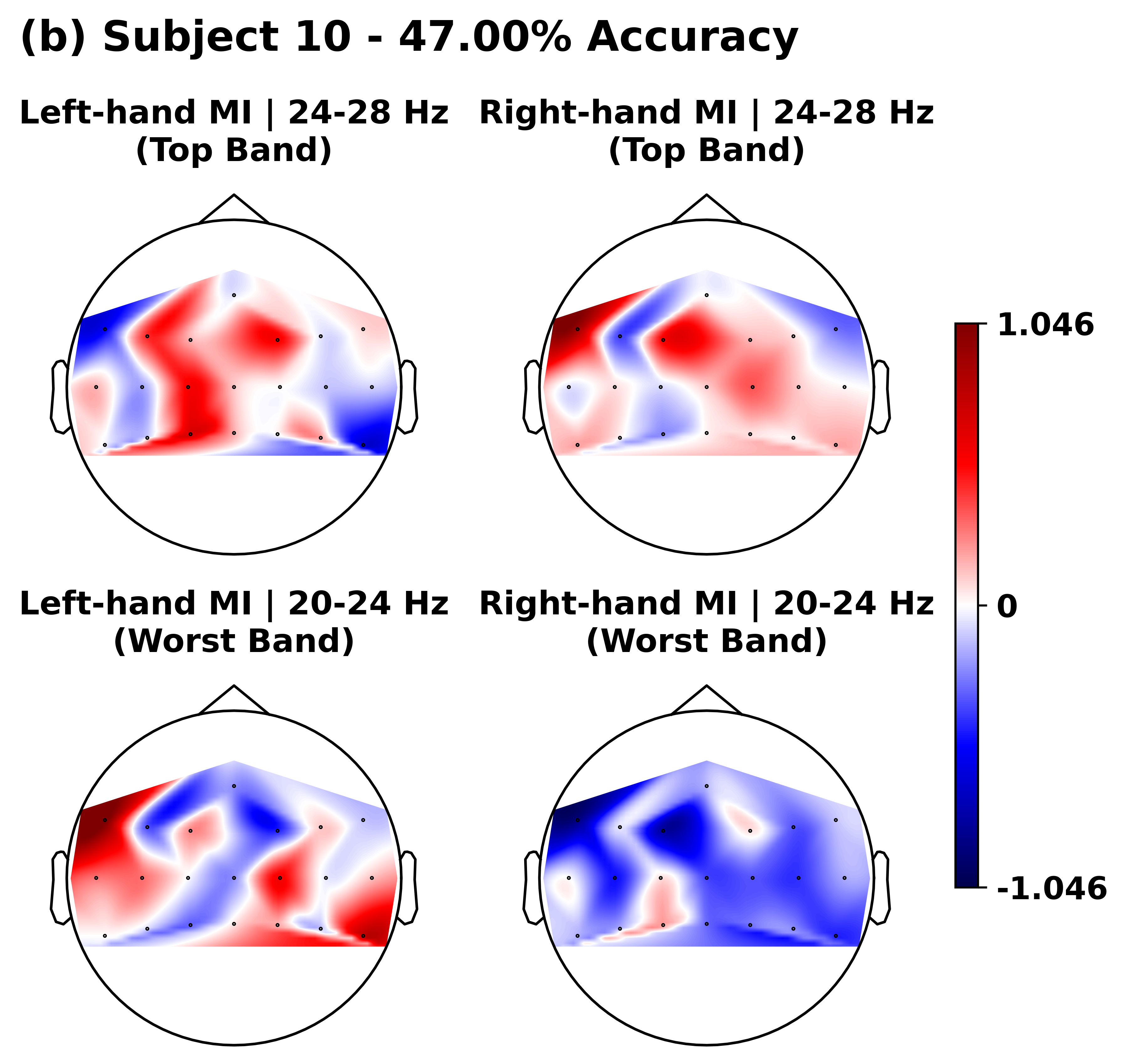}
         \caption{wPLI topomaps for the OpenBMI dataset ($n=54$) during session 2. Panels depict the subjects with the \textbf{(a)} highest (Subject 36) and \textbf{(b)} lowest (Subject 10) mean classification accuracies, evaluated at $K^*=7$ with no MIBIF. Topomaps visualize CSP spatial patterns. For each specified frequency band, a CSP model was fit across all trials to distinguish left- and right-hand motor imagery. Color gradients correspond to the spatial weights of the extreme CSP components (representing the distinct classes) mapped onto 21 selected motor cortex electrode locations. Within each panel, the upper and lower topomaps display the most and least frequently selected frequency bands, respectively. Band selection frequencies were aggregated across the 10 cross-validation folds. The upper topomaps display the most frequently selected band for each subject: 12-16 Hz (selected 100.0\% of the time) for Subject 36, and 24-28 Hz (100.0\%) for Subject 10. The lower topomaps display the least frequently selected band for each subject: 28-32 Hz (30.0\%) for Subject 36, and 20-24 Hz (20.0\%) for Subject 10.}
         \label{fig:topomaps_wpli}
     \end{figure}

In contrast to PLV, the CSP spatial patterns in the bands selected by wPLI appear more spatially distributed and less focal. For the best-performing wPLI subject (\Cref{fig:topomaps_wpli}(a)), the pipeline identified the 12-16 Hz (low-$\beta$ range) as a discriminative feature in 100.0\% of CV folds. The spatial patterns in this band exhibit a distributed structure that appears less dominated by a discrete sensorimotor focus. Notably, the visual difference between the spatial topographies of the top and the worst band (28-32 Hz, selected in only 30.0\% of CV folds) is less pronounced than in the PLV results for the same subject. Combined with the high performance of this subject, this suggests that the discriminative structure of the frequency bands selected by wPLI is more distributed and less dominated by sensorimotor focus and lateralization.

The topomaps for the worst-performing wPLI subject (\Cref{fig:topomaps_wpli}(b)) provide context for the lower classification accuracy observed. Although the pipeline consistently selected the 24–28 Hz band in 100.0\% of CV folds, the resulting CSP patterns do not show a clear task-specific discriminative sensorimotor structure. This suggests weaker class-separating variance, which aligns with the near-chance performance recorded for this subject.

\subsection{Summary of Main Findings}

FC-guided band selection successfully preserves near-baseline classification accuracy with fewer frequency bands, yielding substantial computational savings compared to the traditional nine-band FBCSP approach. PLV enables the most aggressive dimensionality reduction by demonstrating high intra-subject and inter-subject band selection stability and consistently isolating canonical motor imagery bands. Topographic analysis suggests that this performance is achieved by PLV's ability to prioritize frequency bands with focal contralateral spatial patterns that allow for strong class-separating variance. In contrast, wPLI's band selections are much more dispersed across the 4-40 Hz spectrum and correspond to more distributed spatial structures with less sensorimotor focus. Despite this dispersion, wPLI demonstrates superior inter-session stability, successfully maintaining or decreasing the required band thresholds ($K^*$) across recording days. Nonetheless, broader inter-subject variability naturally attenuates dimensionality reduction rates, as evidenced by the larger OpenBMI dataset requiring the retention of more spectral information to maintain baseline equivalence compared to the smaller BCIC IV-2a cohort.

\section{DISCUSSION}

The findings of this study provide a proof of concept that functional connectivity can serve as a principled, subject-specific criterion to reduce the FBCSP filter bank while preserving near-baseline performance. By leveraging the neurophysiological relationships captured by FC metrics, this approach offers a systematic and interpretable alternative to exhaustive or heuristic frequency band selection. In many of the evaluated configurations, this targeted selection preserved near-baseline performance and significantly outperformed random ablation, indicating that the retained bands carry discriminative information beyond that of an arbitrary subset of the same size.

Among the metrics evaluated, PLV emerged as the strongest candidate for aggressive dimensionality reduction. As evidenced by the analysis of band selection stability, PLV consistently isolated bands within the $\mu$ and low-$\beta$ ranges, which are canonically associated with motor imagery \cite{pfurtscheller_motor_1997,neuper_event-related_2001,jeon_event-related_2011}. This concentrated selection highlights physiological interpretability as an advantage of the proposed method, and topographic analysis of the best-performing subject showed that PLV prioritized bands with focal contralateral CSP patterns. However, PLV's consistency in selecting the 12-16 Hz (low-$\beta$) and 8-12 Hz ($\mu$) bands may be substantially influenced by volume conduction. Because the evaluated channel pairs are physically adjacent on the scalp, the strong localized event-related desynchronization of the $\mu$ and $\beta$ rhythms likely spreads to the midline electrodes. Since PLV does not penalize zero-lag phase synchrony, it may be prioritizing these bands based on localized signal spread rather than long-range network communication. These two observations are not mutually exclusive: the bands selected by PLV remain physiologically interpretable, but the mechanism driving their selection may be closer to lateralized spectral power than to inter-regional phase coupling, a distinction that can be tested directly by comparing FC-based ranking against ranking based on band-power lateralization. Conversely, wPLI yielded more dispersed band selection, choosing bands with distributed CSP spatial patterns, while demonstrating greater inter-session consistency in terms of $K^*$. Because wPLI suppresses zero-lag interactions to mitigate volume conduction, its dispersed selection pattern may reflect more subtle, time-lagged functional connectivity, which may explain why it generally requires a higher $K^*$ to achieve near-baseline performance. Alternatively, the higher $K^*$ may partly reflect the greater variance of single-trial wPLI estimates over short epochs, which would reduce the reliability of its band rankings independently of the underlying connectivity. Thus, while wPLI may not facilitate the same degree of dimensionality reduction as PLV, it may capture connectivity patterns that are more stable across recording sessions, although this should be confirmed under cross-session evaluation.

The generalizability of these findings is contextualized by the differences observed between the two datasets. In the smaller BCIC IV-2a dataset ($n=9$), the PLV-guided pipeline achieved baseline equivalence with as few as 2 to 3 frequency bands. However, in the larger OpenBMI dataset ($n=54$), the minimum number of bands ($K^*$) required for near-baseline performance with PLV increased to 5 or 6. This likely reflects greater inter-subject variability in the larger cohort, as well as differences in channel montage, epoch window, and recording setup between the two datasets. Nonetheless, the FC-guided framework remains practically relevant even when the degree of dimensionality reduction is less aggressive. For instance, reaching baseline performance with $K^*=5$ or $K^*=6$ on OpenBMI still corresponds to a 44.4\% or 33.3\% reduction in CSP training fits, respectively. FC-guided band selection therefore offers a principled way to reduce the number of spatial filters that must be trained without sacrificing classification performance, which may be valuable in MI-BCI settings where computational overhead is constrained, such as embedded or online systems.

While these findings demonstrate the feasibility of FC-guided band selection, several limitations must be acknowledged. First, the contribution of this work lies in the band selection criterion rather than in a new decoding architecture, as it integrates established FC metrics into a well-known FBCSP pipeline. Second, the success and degree of band reduction are highly dependent on the specific FC metric utilized, and the values of $K^*$ and the significance over random ablation varied across FC metrics, datasets, and recording sessions. Third, the reported savings count only the reduction in CSP fits and do not account for the cost of computing FC across all candidate bands, which must be weighed against those savings in an end-to-end pipeline. Fourth, the inter-subject variability, which is particularly evident in the OpenBMI dataset, indicates that while FC-guided selection is subject-specific, it cannot entirely overcome ``BCI inefficiency'' \cite{lee_eeg_2019} or the inherent performance ceilings observed in certain users. Finally, the current methodology was developed and validated exclusively within an FBCSP framework, which has previously been shown to be less robust to channel-specific noise than EEGNet \cite{Nagarajan2021InvestigationInterfaces}. Whether FC-guided band selection can be effectively extended to contemporary motor imagery decoders, such as deep learning architectures \cite{padfield_eeg-based_2019, tayeb_validating_2019}, and in particular to filter-bank deep decoders, therefore remains an open question and a critical direction for future research.

Beyond the choice of decoder, the spectral dimension is only one axis along which MI-BCI pipelines can be pruned. Principled selection has previously been shown to reduce the channel dimension, with relevance-guided selection removing 61\% of channels without a significant loss in subject-independent accuracy \cite{Nagarajan2023Relevance-basedInterface}. The present results suggest that an analogous, though more dataset-dependent, reduction is achievable in the spectral domain, making joint selection over channels and frequency bands a natural extension of this work.

\section{CONCLUSIONS}

This study establishes a successful proof of concept for subject-specific frequency band selection within MI-BCI FBCSP pipelines using band-specific phase-based functional connectivity. The proposed framework demonstrates that near-baseline classification accuracy can be preserved while achieving a 22.2 to 77.8\% reduction in the required CSP spatial filter fits across the evaluated configurations. While the PLV metric enables the most aggressive dimensionality reduction by consistently selecting bands in the $\mu$ and low-$\beta$ ranges with strong lateralized CSP patterns, the wPLI metric provides superior inter-session robustness by mitigating the influence of volume conduction. Although the degree of dimensionality reduction is attenuated by the high inter-subject variability inherent in larger cohorts, the static FC-guided approach significantly outperformed random band ablation in several configurations. These findings suggest that static FC-based band selection offers a systematic alternative to heuristic filter bank designs, providing a potential avenue for less computationally expensive and more interpretable MI-BCI pipelines. Future work should explore the generalizability of these FC-guided criteria within other contemporary decoding architectures, such as deep learning and geometry-based models.

\FloatBarrier

\section*{APPENDIX}

\subsection{Supplementary Statistical Tables}
\input{tables/plv_posthoc}

\FloatBarrier






\FloatBarrier

\bibliographystyle{ieeetr}
\bibliography{references}





\end{document}

%% file: alg_pipeline.tex
\begin{algorithm}[htbp]
\caption{Proposed FC-Based Band Selection and Classification Pipeline}
\label{alg:proposed_method}
\begin{algorithmic}[1]
\renewcommand{\algorithmicrequire}{\textbf{Input:}}
\renewcommand{\algorithmicensure}{\textbf{Output:}}

\Require Raw EEG data $\mathcal{D}$, Set of bands $B$, Number of selected bands $K$, Boolean MIBIF
\Ensure Mean cross-validation accuracy $\overline{Acc}$

\State $\mathcal{D}_{pre} \gets \text{Preprocess}(\mathcal{D})$
\State $Acc_{list} \gets [ ]$

\For{$fold \gets 1$ \textbf{to} $10$}
    \State $\mathcal{D}_{train}, \mathcal{D}_{test} \gets \text{Split}(\mathcal{D}_{pre}, fold)$
    
    \For{\textbf{each} band $b \in B$}
        \State Compute FC on $\mathcal{D}_{train}$
        \State Calculate $C_\text{L}$ and $C_\text{R}$ for each trial
        \State Compute coupling difference $\Delta C = C_\text{L} - C_\text{R}$
        \State Calculate $M_\text{L}, M_\text{R}$, and $S_\text{pooled}$ from $\Delta C$ distributions
        \State Compute modified Cohen's $d$ for band $b$
    \EndFor
    
    \State $B_{ranked} \gets \text{Sort } B \text{ by descending order of } d$
    \State $B^* \gets \text{Select top } K \text{ bands from } B_{ranked}$
    
    \State $\mathbf{F}_{train}, \mathbf{F}_{test} \gets \text{Extract FBCSP features using } B^*$
    
    \If{MIBIF \textbf{is} True}
        \State Apply MIBIF selection to $\mathbf{F}_{train}$ and $\mathbf{F}_{test}$
    \EndIf
    
    \State $\mathcal{M} \gets \text{Train SVM}(\mathbf{F}_{train}, \text{labels}_{train})$
    \State $Acc_{fold} \gets \text{Evaluate}(\mathcal{M}, \mathbf{F}_{test})$
    \State Append $Acc_{fold}$ to $Acc_{list}$
\EndFor

\State $\overline{Acc} \gets \text{Mean}(Acc_{list})$
\State \Return $\overline{Acc}$

\end{algorithmic}
\end{algorithm}

%% file: tables/baseline_accuracy.tex
\begin{tabular}{lcc}
    \toprule
     & \makecell{BCIC IV-2a \\ ($n=9$)} & \makecell{OpenBMI \\ ($n=54$)} \\
    \midrule
    MIBIF (4 features) & 84.10 $\pm$ 16.59 & 68.60 $\pm$ 17.61 \\
    No MIBIF      & 85.41 $\pm$ 14.59 & 67.61 $\pm$ 17.68 \\
    \bottomrule
\end{tabular}

%% file: tables/mink_bcic_table.tex
\begin{tabular}{llccc}
    \toprule
    FC Metric & Feature Selection & $K^*$ \\
    \midrule
    wPLI & MIBIF$^\dagger$ & 7 \\
     & No MIBIF & 7 \\ \addlinespace
    PLV & MIBIF$^\dagger$ & \textbf{2\rlap{$^*$}} \\
     & No MIBIF & 3\rlap{$^*$} \\ \addlinespace
    PLI & MIBIF$^\dagger$ & 6 \\
    & No MIBIF & 8 \\
    \bottomrule
\end{tabular}

%% file: tables/mink_openbmi_table.tex
\setlength{\defaultaddspace}{0.4em}
\begin{tabular}{llccc}
    \toprule
    \makecell[l]{FC \\ Metric} & \makecell[l]{Feature \\ Selection} & \makecell{Session 1 \\ $K^*$} & \makecell{Session 2 \\ $K^*$} & \makecell{Combined \\ $K^*$} \\
    \midrule
    wPLI & MIBIF$^\dagger$ & 7 & \textbf{6\rlap{$^*$}} & 6\rlap{$^{**}$} \\
     & No MIBIF & 7 & 7 & 7 \\ \addlinespace
    PLV & MIBIF$^\dagger$ & 6\rlap{$^{**}$} & 7 & 6\rlap{$^*$} \\
     & No MIBIF & \textbf{5\rlap{$^{**}$}} & 7\rlap{$^*$} & \textbf{5\rlap{$^{**}$}} \\ \addlinespace
    PLI & MIBIF$^\dagger$ & 7 & 7\rlap{$^*$} & 7\rlap{$^*$} \\
    & No MIBIF & 6\rlap{$^*$} & 8 & 6\rlap{$^*$} \\
    \bottomrule
\end{tabular}

%% file: tables/bcic_wilcoxon_ablation.tex
\resizebox{\columnwidth}{!}{%
\setlength{\defaultaddspace}{0.4em}
\begin{tabular}{llcccccc}
    \toprule
    \makecell[l]{FC \\ Metric} & \makecell[l]{Feature \\ Selection} & $K^*$ & \makecell{FC Mean \\ Acc (\%)} & \makecell{Random Mean \\ Acc (\%)} & \makecell{$\Delta$ FC vs \\ Random (\%)} & $W$ & $p_{\text{adj}}$ \\
    \midrule
    wPLI & MIBIF$^\dagger$ & 7 & 83.58 & 83.30 & 0.28 & 22.0 & 0.5449 \\
     & No MIBIF & 7 & 85.08 & 83.74 & 1.34 & 36.0 & 0.1934 \\ \addlinespace
    PLV & MIBIF$^\dagger$ & 2 & 83.84 & 73.21 & 10.63 & 44.0 & 0.0234\rlap{$^*$} \\
     & No MIBIF & 3 & 84.84 & 76.45 & 8.39 & 44.0 & 0.0234\rlap{$^*$} \\ \addlinespace
    PLI & MIBIF$^\dagger$ & 6 & 83.90 & 82.38 & 1.52 & 39.0 & 0.1094 \\
     & No MIBIF & 8 & 85.42 & 84.75 & 0.67 & 32.0 & 0.3008 \\
    \bottomrule
\end{tabular}
}

%% file: tables/openbmi_wilcoxon_ablation.tex
\resizebox{\columnwidth}{!}{%
\setlength{\defaultaddspace}{0.4em}
\begin{tabular}{llcccccc}
    \toprule
    \makecell[l]{FC \\ Metric} & \makecell[l]{Feature \\ Selection} & $K^*$ & \makecell{FC Mean \\ Acc (\%)} & \makecell{Random Mean \\ Acc (\%)} & \makecell{$\Delta$ FC vs \\ Random (\%)} & $W$ & $p_{\text{adj}}$ \\
    \midrule
    \multicolumn{8}{c}{\textbf{Session 1}} \\
    \midrule
    wPLI & MIBIF$^\dagger$ & 7 & 67.09 & 66.17 & 0.92 & 1023.0 & 0.0550 \\
     & No MIBIF & 7 & 65.96 & 65.10 & 0.86 & 995.5 & 0.0595 \\ \addlinespace
    PLV & MIBIF$^\dagger$ & 6 & 67.80 & 65.19 & 2.61 & 1243.0 & 0.0001\rlap{$^{**}$} \\
     & No MIBIF & 5 & 66.57 & 63.30 & 3.27 & 1231.0 & 0.0002\rlap{$^{**}$} \\ \addlinespace
    PLI & MIBIF$^\dagger$ & 7 & 67.19 & 66.17 & 1.02 & 964.5 & 0.0595 \\
     & No MIBIF & 6 & 65.74 & 64.31 & 1.43 & 1076.5 & 0.0222\rlap{$^*$} \\
    \midrule
    \multicolumn{8}{c}{\textbf{Session 2}} \\
    \midrule
    wPLI & MIBIF$^\dagger$ & 6 & 68.44 & 66.64 & 1.80 & 1099.5 & 0.0127\rlap{$^*$} \\
     & No MIBIF & 7 & 68.04 & 67.10 & 0.94 & 939.0 & 0.0595 \\ \addlinespace
    PLV & MIBIF$^\dagger$ & 7 & 68.59 & 67.69 & 0.90 & 1005.0 & 0.0595 \\
     & No MIBIF & 7 & 68.09 & 67.10 & 0.99 & 1014.0 & 0.0370\rlap{$^*$} \\ \addlinespace
    PLI & MIBIF$^\dagger$ & 7 & 68.67 & 67.69 & 0.98 & 1033.5 & 0.0489\rlap{$^*$} \\
     & No MIBIF & 8 & 67.89 & 68.10 & -0.21 & 658.0 & 0.6946 \\
    \midrule
    \multicolumn{8}{c}{\textbf{Combined Sessions}} \\
    \midrule
    wPLI & MIBIF$^\dagger$ & 6 & 67.40 & 65.92 & 1.48 & 1194.0 & 0.0008\rlap{$^{**}$} \\
     & No MIBIF & 7 & 67.00 & 66.10 & 0.90 & 1017.0 & 0.0550 \\ \addlinespace
    PLV & MIBIF$^\dagger$ & 6 & 67.83 & 65.92 & 1.91 & 1176.0 & 0.0013\rlap{$^*$} \\
     & No MIBIF & 5 & 66.85 & 64.15 & 2.70 & 1195.5 & 0.0008\rlap{$^{**}$} \\ \addlinespace
    PLI & MIBIF$^\dagger$ & 7 & 67.93 & 66.93 & 1.00 & 1072.5 & 0.0225\rlap{$^*$} \\
     & No MIBIF & 6 & 66.69 & 65.28 & 1.41 & 1147.0 & 0.0032\rlap{$^*$} \\
    \bottomrule
\end{tabular}
}

%% file: tables/compute_savings.tex
\resizebox{\columnwidth}{!}{%
\setlength{\defaultaddspace}{0.4em}
\begin{tabular}{lllccc}
    \toprule
    Dataset & \makecell[l]{FC \\ Metric} & \makecell[l]{Feature \\ Selection} & \makecell{Baseline \\ Bands} & $K^*$ & \makecell{Compute \\ Savings} \\
    \midrule
    \multirow{2}{*}{\makecell[l]{BCIC IV-2a \\ ($n=9$)}} 
    & PLV & MIBIF$^\dagger$ & 9 & 2 & 77.8\% \\
    & & No MIBIF & 9 & 3 & 66.7\% \\
    \midrule
    \multirow{5}{*}{\makecell[l]{OpenBMI \\ ($n=54$)}} 
    & wPLI & MIBIF$^\dagger$ & 9 & 6 & 33.3\% \\ \addlinespace
    & PLV & MIBIF$^\dagger$ & 9 & 6 & 33.3\% \\
    & & No MIBIF & 9 & 5 & 44.4\% \\ \addlinespace
    & PLI & MIBIF$^\dagger$ & 9 & 7 & 22.2\% \\
    & & No MIBIF & 9 & 6 & 33.3\% \\
    \bottomrule
\end{tabular}
}

%% file: tables/plv_posthoc.tex
\begin{table}[htbp]
\centering
\caption{Post-hoc pairwise comparisons for PLV band selection frequencies}
\label{tab:plv_posthoc}
\begin{tabular}{lcc}
\toprule
Comparison (Hz) & Adjusted $p$-value & Significance \\
\midrule
12--16 vs 16--20 & \textbf{$<$ 0.001} & \textbf{**} \\
12--16 vs 20--24 & \textbf{0.002} & \textbf{*} \\
12--16 vs 24--28 & \textbf{0.002} & \textbf{*} \\
12--16 vs 28--32 & \textbf{$<$ 0.001} & \textbf{**} \\
12--16 vs 32--36 & \textbf{$<$ 0.001} & \textbf{**} \\
12--16 vs 36--40 & \textbf{$<$ 0.001} & \textbf{**} \\
12--16 vs 4--8   & \textbf{$<$ 0.001} & \textbf{**} \\
12--16 vs 8--12  & \textbf{0.015} & \textbf{*} \\
16--20 vs 20--24 & 0.315 & \\
16--20 vs 24--28 & 0.237 & \\
16--20 vs 28--32 & 0.573 & \\
16--20 vs 32--36 & 0.399 & \\
16--20 vs 36--40 & 0.573 & \\
16--20 vs 4--8   & 0.130 & \\
16--20 vs 8--12  & \textbf{0.015} & \textbf{*} \\
20--24 vs 24--28 & 0.496 & \\
20--24 vs 28--32 & 0.272 & \\
20--24 vs 32--36 & 0.073 & \\
20--24 vs 36--40 & 0.176 & \\
20--24 vs 4--8   & \textbf{0.019} & \textbf{*} \\
20--24 vs 8--12  & 0.176 & \\
24--28 vs 28--32 & \textbf{0.021} & \textbf{*} \\
24--28 vs 32--36 & \textbf{0.015} & \textbf{*} \\
24--28 vs 36--40 & 0.077 & \\
24--28 vs 4--8   & \textbf{0.008} & \textbf{*} \\
24--28 vs 8--12  & 0.320 & \\
28--32 vs 32--36 & 0.320 & \\
28--32 vs 36--40 & 0.676 & \\
28--32 vs 4--8   & 0.066 & \\
28--32 vs 8--12  & \textbf{0.017} & \textbf{*} \\
32--36 vs 36--40 & 0.833 & \\
32--36 vs 4--8   & 0.237 & \\
32--36 vs 8--12  & \textbf{0.013} & \textbf{*} \\
36--40 vs 4--8   & 0.215 & \\
36--40 vs 8--12  & \textbf{0.015} & \textbf{*} \\
4--8 vs 8--12    & \textbf{0.002} & \textbf{*} \\
\bottomrule
\end{tabular}

\vspace{2pt}
        {\justifying\footnotesize\noindent Post-hoc comparisons for FC-guided band selection frequencies with session 2 of the OpenBMI dataset ($n=54$). A Friedman test revealed significant differences across bands ($\chi^2(8) = 55.770$, $p < 0.001$). Pairwise comparisons were assessed using Wilcoxon signed-rank tests with FDR correction (Benjamini-Hochberg). Significant results ($p_{\text{adj}} < 0.05$) are bolded. The * and ** denote statistical significance, where *: $p_{\text{adj}} < 0.05$ and **: $p_{\text{adj}} < 0.001$.\par} 
\end{table}

%% file: references.bib
@article{ang_eeg-based_2017,
    author = {Ang, Kai Keng and Guan, Cuntai},
    title = {{EEG}-Based Strategies to Detect Motor Imagery for Control and Rehabilitation},
    journal = {IEEE Transactions on Neural Systems and Rehabilitation Engineering},
    volume = {25},
    number = {4},
    pages = {392--401},
    year = {2017},
    month = apr,
    doi = {10.1109/TNSRE.2016.2646763}
}

@article{tangermann_review_2012,
    author = {Tangermann, Michael and others},
    title = {Review of the {BCI} Competition {IV}},
    journal = {Frontiers in Neuroscience},
    volume = {6},
    pages = {55},
    year = {2012},
    month = jul,
    doi = {10.3389/fnins.2012.00055}
}

@article{lee_eeg_2019,
    author = {Lee, Min-Ho and others},
    title = {{EEG} dataset and {OpenBMI} toolbox for three {BCI} paradigms: an investigation into {BCI} illiteracy},
    journal = {GigaScience},
    volume = {8},
    number = {5},
    pages = {giz002},
    year = {2019},
    month = may,
    doi = {10.1093/gigascience/giz002}
}

@inproceedings{ang_filter_2008,
    author = {Ang, Kai Keng and Chin, Zheng Yang and Zhang, Haihong and Guan, Cuntai},
    title = {Filter Bank Common Spatial Pattern ({FBCSP}) in Brain-Computer Interface},
    booktitle = {2008 {IEEE} International Joint Conference on Neural Networks ({IEEE} World Congress on Computational Intelligence)},
    pages = {2390--2397},
    year = {2008},
    month = jun,
    doi = {10.1109/IJCNN.2008.4634130}
}

@misc{larson_mne-python_2024,
    author = {Larson, Eric and others},
    title = {{MNE-Python}},
    publisher = {Zenodo},
    year = {2024},
    month = aug,
    doi = {10.5281/zenodo.13340330}
}

@article{gramfort_meg_2013,
    author = {Gramfort, Alexandre and others},
    title = {{MEG} and {EEG} data analysis with {MNE-Python}},
    journal = {Frontiers in Neuroscience},
    volume = {7},
    pages = {267},
    year = {2013},
    month = dec,
    doi = {10.3389/fnins.2013.00267}
}

@misc{li_mne-connectivity_2026,
    author = {Li, Adam and others},
    title = {mne-connectivity},
    publisher = {Zenodo},
    year = {2026},
    month = mar,
    doi = {10.5281/zenodo.10278399}
}

@article{vinck_improved_2011,
    author = {Vinck, Martin and Oostenveld, Robert and van Wingerden, Marijn and Battaglia, Franscesco and Pennartz, Cyriel M. A.},
    title = {An improved index of phase-synchronization for electrophysiological data in the presence of volume-conduction, noise and sample-size bias},
    journal = {NeuroImage},
    volume = {55},
    number = {4},
    pages = {1548--1565},
    year = {2011},
    month = apr,
    doi = {10.1016/j.neuroimage.2011.01.055}
}

@inproceedings{wang_phase_2006,
    author = {Wang, Yijun and Hong, Bo and Gao, Xiaorong and Gao, Shangkai},
    title = {Phase Synchrony Measurement in Motor Cortex for Classifying Single-trial {EEG} during Motor Imagery},
    booktitle = {2006 International Conference of the {IEEE} Engineering in Medicine and Biology Society},
    pages = {75--78},
    year = {2006},
    month = aug,
    doi = {10.1109/IEMBS.2006.259673}
}

@article{wang_diverse_2020,
    author = {Wang, Hongtao and Xu, Xiaowei and Yin, Yuhang and Li, Zhenghao and Zhang, Zhilin},
    title = {Diverse Feature Blend Based on Filter-Bank Common Spatial Pattern and Brain Functional Connectivity for Multiple Motor Imagery Detection},
    journal = {IEEE Access},
    volume = {8},
    pages = {168331--168341},
    year = {2020},
    month = aug,
    doi = {10.1109/ACCESS.2020.3018962}
}

@article{darvishi_eeg-driven_2025,
    author = {Darvishi, Hamidreza and Mohammadi, Ahmadreza and Maghami, Mohammad Hossein and Sadeghi, Meysam and Sawan, Mohamad},
    title = {{EEG}-Driven Arm Movement Decoding: Combining Connectivity and Amplitude Features for Enhanced Brain-Computer Interface Performance},
    journal = {Bioengineering},
    volume = {12},
    number = {6},
    pages = {614},
    year = {2025},
    month = jun,
    doi = {10.3390/bioengineering12060614}
}

@inproceedings{gonuguntla_phase_2013,
    author = {Gonuguntla, V. and Wang, Y. and Veluvolu, K. C.},
    title = {Phase synchrony in subject-specific reactive band of {EEG} for classification of motor imagery tasks},
    booktitle = {2013 35th Annual International Conference of the {IEEE} Engineering in Medicine and Biology Society ({EMBC})},
    pages = {2784--2787},
    year = {2013},
    month = jul,
    doi = {10.1109/EMBC.2013.6610118}
}

@article{leeuwis_functional_2021,
    author = {Leeuwis, Nikki and Yoon, Sue and Alimardani, Maryam},
    title = {Functional Connectivity Analysis in Motor-Imagery Brain Computer Interfaces},
    journal = {Frontiers in Human Neuroscience},
    volume = {15},
    year = {2021},
    month = oct,
    doi = {10.3389/fnhum.2021.732946}
}

@article{stam_phase_2007,
    author = {Stam, Cornelis J. and Nolte, Guido and Daffertshofer, Andreas},
    title = {Phase lag index: Assessment of functional connectivity from multi channel {EEG} and {MEG} with diminished bias from common sources},
    journal = {Human Brain Mapping},
    volume = {28},
    number = {11},
    pages = {1178--1193},
    year = {2007},
    month = nov,
    doi = {10.1002/hbm.20346}
}

@article{scikit-learn,
    author = {Pedregosa, F. and others},
    title = {Scikit-learn: Machine Learning in {Python}},
    journal = {Journal of Machine Learning Research},
    volume = {12},
    pages = {2825--2830},
    year = {2011}
}

@article{pfurtscheller_motor_1997,
    author = {Pfurtscheller, Gert and Neuper, Christa},
    title = {Motor imagery activates primary sensorimotor area in humans},
    journal = {Neuroscience Letters},
    volume = {239},
    number = {2-3},
    pages = {65--68},
    year = {1997},
    month = dec,
    doi = {10.1016/S0304-3940(97)00889-6}
}

@article{neuper_event-related_2001,
    author = {Neuper, C. and Pfurtscheller, G.},
    title = {Event-related dynamics of cortical rhythms: frequency-specific features and functional correlates},
    journal = {International Journal of Psychophysiology},
    volume = {43},
    number = {1},
    pages = {41--58},
    year = {2001},
    month = dec,
    doi = {10.1016/S0167-8760(01)00178-7}
}

@article{jeon_event-related_2011,
    author = {Jeon, Yongwoong and Nam, Chang S. and Kim, Young-Joo and Whang, Min Cheol},
    title = {Event-related (De)synchronization ({ERD}/{ERS}) during motor imagery tasks: Implications for brain-computer interfaces},
    journal = {International Journal of Industrial Ergonomics},
    volume = {41},
    number = {5},
    pages = {428--436},
    year = {2011},
    month = sep,
    doi = {10.1016/j.ergon.2011.03.005}
}

@article{padfield_eeg-based_2019,
    author = {Padfield, Natasha and Zabalza, Jaime and Zhao, Huimin and Masero, Valentin and Ren, Jinchang},
    title = {{EEG}-Based Brain-Computer Interfaces Using Motor-Imagery: Techniques and Challenges},
    journal = {Sensors},
    volume = {19},
    number = {6},
    pages = {1423},
    year = {2019},
    month = mar,
    doi = {10.3390/s19061423}
}

@article{tayeb_validating_2019,
    author = {Tayeb, Zied and others},
    title = {Validating Deep Neural Networks for Online Decoding of Motor Imagery Movements from {EEG} Signals},
    journal = {Sensors},
    volume = {19},
    number = {1},
    pages = {210},
    year = {2019},
    month = jan,
    doi = {10.3390/s19010210}
}

@article{wolpaw_braincomputer_2002,
    author = {Wolpaw, Jonathan R. and Birbaumer, Niels and McFarland, Dennis J. and Pfurtscheller, Gert and Vaughan, Theresa M.},
    title = {Brain-computer interfaces for communication and control},
    journal = {Clinical Neurophysiology},
    volume = {113},
    number = {6},
    pages = {767--791},
    year = {2002},
    month = jun,
    doi = {10.1016/S1388-2457(02)00057-3}
}

@article{nicolas-alonso_brain_2012,
    author = {Nicolas-Alonso, Luis Fernando and Gomez-Gil, Jaime},
    title = {Brain Computer Interfaces, a Review},
    journal = {Sensors (Basel, Switzerland)},
    volume = {12},
    number = {2},
    pages = {1211--1279},
    year = {2012},
    month = jan,
    doi = {10.3390/s120201211}
}

@article{mcfarland_eeg-based_2017,
    author = {McFarland, D. J. and Wolpaw, J. R.},
    title = {{EEG}-based brain-computer interfaces},
    journal = {Current Opinion in Biomedical Engineering},
    volume = {4},
    pages = {194--200},
    year = {2017},
    month = dec,
    doi = {10.1016/j.cobme.2017.11.004}
}

@article{shih_brain-computer_2012,
    author = {Shih, Jerry J. and Krusienski, Dean J. and Wolpaw, Jonathan R.},
    title = {Brain-Computer Interfaces in Medicine},
    journal = {Mayo Clinic Proceedings},
    volume = {87},
    number = {3},
    pages = {268--279},
    year = {2012},
    month = mar,
    doi = {10.1016/j.mayocp.2011.12.008}
}

@article{cohen_where_2017,
    author = {Cohen, Michael X.},
    title = {Where Does {EEG} Come From and What Does It Mean?},
    journal = {Trends in Neurosciences},
    volume = {40},
    number = {4},
    pages = {208--218},
    year = {2017},
    month = apr,
    doi = {10.1016/j.tins.2017.02.004}
}

@article{mcfarland_brain-computer_2005,
    author = {McFarland, Dennis J. and Sarnacki, William A. and Vaughan, Theresa M. and Wolpaw, Jonathan R.},
    title = {Brain-computer interface ({BCI}) operation: signal and noise during early training sessions},
    journal = {Clinical Neurophysiology},
    volume = {116},
    number = {1},
    pages = {56--62},
    year = {2005},
    month = jan,
    doi = {10.1016/j.clinph.2004.07.004}
}

@article{pelgrims_contribution_2010,
    author = {Pelgrims, Barbara and Michaux, Nicolas and Olivier, Etienne and Andres, Michael},
    title = {Contribution of the primary motor cortex to motor imagery: A subthreshold {TMS} study},
    journal = {Human Brain Mapping},
    volume = {32},
    number = {9},
    pages = {1471--1482},
    year = {2010},
    month = nov,
    doi = {10.1002/hbm.21121}
}

@article{barone_understanding_2021,
    author = {Barone, Jacopo and Rossiter, Holly E.},
    title = {Understanding the Role of Sensorimotor Beta Oscillations},
    journal = {Frontiers in Systems Neuroscience},
    volume = {15},
    year = {2021},
    month = may,
    doi = {10.3389/fnsys.2021.655886}
}

@article{shanechi_brainmachine_2019,
    author = {Shanechi, Maryam M.},
    title = {Brain-machine interfaces from motor to mood},
    journal = {Nature Neuroscience},
    volume = {22},
    number = {10},
    pages = {1554--1564},
    year = {2019},
    month = oct,
    doi = {10.1038/s41593-019-0488-y}
}

@article{muller-gerking_designing_1999,
    author = {M{\"u}ller-Gerking, Johannes and Pfurtscheller, Gert and Flyvbjerg, Henrik},
    title = {Designing optimal spatial filters for single-trial {EEG} classification in a movement task},
    journal = {Clinical Neurophysiology},
    volume = {110},
    number = {5},
    pages = {787--798},
    year = {1999},
    month = may,
    doi = {10.1016/S1388-2457(98)00038-8}
}

@article{pfurtscheller_mu_2006,
    author = {Pfurtscheller, G. and Brunner, C. and Schl{\"o}gl, A. and Lopes da Silva, F. H.},
    title = {Mu rhythm (de)synchronization and {EEG} single-trial classification of different motor imagery tasks},
    journal = {NeuroImage},
    volume = {31},
    number = {1},
    pages = {153--159},
    year = {2006},
    month = may,
    doi = {10.1016/j.neuroimage.2005.12.003}
}

@article{lotte_review_2018,
    author = {Lotte, F. and others},
    title = {A review of classification algorithms for {EEG}-based brain-computer interfaces: a 10 year update},
    journal = {Journal of Neural Engineering},
    volume = {15},
    number = {3},
    pages = {031005},
    year = {2018},
    month = apr,
    doi = {10.1088/1741-2552/aab2f2}
}

@article{miladinovic_performance_2020,
    author = {Miladinovi{\'c}, Aleksandar and others},
    title = {Performance of {EEG} Motor-Imagery based spatial filtering methods: A {BCI} study on Stroke patients},
    journal = {Procedia Computer Science},
    volume = {176},
    pages = {2840--2848},
    year = {2020},
    month = jan,
    doi = {10.1016/j.procs.2020.09.270}
}

@article{ang_filter_2012,
    author = {Ang, Kai Keng and Chin, Zheng Yang and Wang, Chuanchu and Guan, Cuntai and Zhang, Haihong},
    title = {Filter Bank Common Spatial Pattern Algorithm on {BCI} Competition {IV} Datasets 2a and 2b},
    journal = {Frontiers in Neuroscience},
    volume = {6},
    pages = {39},
    year = {2012},
    month = mar,
    doi = {10.3389/fnins.2012.00039}
}

@article{marzetti_brain_2019,
    author = {Marzetti, Laura and Basti, Alessio and Chella, Federico and D'Andrea, Antea and Syrj{\"a}l{\"a}, Jaakko and Pizzella, Vittorio},
    title = {Brain Functional Connectivity Through Phase Coupling of Neuronal Oscillations: A Perspective From Magnetoencephalography},
    journal = {Frontiers in Neuroscience},
    volume = {13},
    year = {2019},
    month = sep,
    doi = {10.3389/fnins.2019.00964}
}

@article{daly_brain_2012,
    author = {Daly, Ian and Nasuto, Slawomir J. and Warwick, Kevin},
    title = {Brain computer interface control via functional connectivity dynamics},
    journal = {Pattern Recognition},
    volume = {45},
    number = {6},
    pages = {2123--2136},
    year = {2012},
    month = jun,
    doi = {10.1016/j.patcog.2011.04.034}
}

@article{lachaux_measuring_1999,
    author = {Lachaux, Jean-Philippe and Rodriguez, Eugenio and Martinerie, Jacques and Varela, Francisco J.},
    title = {Measuring phase synchrony in brain signals},
    journal = {Human Brain Mapping},
    volume = {8},
    number = {4},
    pages = {194--208},
    year = {1999},
    doi = {10.1002/(SICI)1097-0193(1999)8:4<194::AID-HBM4>3.0.CO;2-C}
}

@article{bruna_phase_2018,
    author = {Bru{\~n}a, Ricardo and Maest{\'u}, Fernando and Pereda, Ernesto},
    title = {Phase locking value revisited: teaching new tricks to an old dog},
    journal = {Journal of Neural Engineering},
    volume = {15},
    number = {5},
    pages = {056011},
    year = {2018},
    month = jul,
    doi = {10.1088/1741-2552/aacfe4}
}

@article{li_common_2019,
    author = {Li, X. and Fan, H. and Wang, H. and Wang, L.},
    title = {Common spatial patterns combined with phase synchronization information for classification of {EEG} signals},
    journal = {Biomedical Signal Processing and Control},
    volume = {52},
    pages = {248--256},
    year = {2019},
    month = jul,
    doi = {10.1016/j.bspc.2019.04.034}
}

@article{siviero_functional_2023,
    author = {Siviero, Ilaria and Menegaz, Gloria and Storti, Silvia Francesca},
    title = {Functional Connectivity and Feature Fusion Enhance Multiclass Motor-Imagery Brain-Computer Interface Performance},
    journal = {Sensors},
    volume = {23},
    number = {17},
    pages = {7520},
    year = {2023},
    month = jan,
    doi = {10.3390/s23177520}
}

@article{ai_feature_2019,
    author = {Ai, Qingsong and others},
    title = {Feature extraction of four-class motor imagery {EEG} signals based on functional brain network},
    journal = {Journal of Neural Engineering},
    volume = {16},
    number = {2},
    pages = {026032},
    year = {2019},
    month = feb,
    doi = {10.1088/1741-2552/ab0328}
}

@article{haufe_interpretation_2014,
    author = {Haufe, Stefan and Meinecke, Frank and G{\"{o}}rgen, Kai and D{\"{a}}hne, Sven and Haynes, John-Dylan and Blankertz, Benjamin and Bie{\ss}mann, Felix},
    title = {On the interpretation of weight vectors of linear models in multivariate neuroimaging},
    journal = {NeuroImage},
    volume = {87},
    pages = {96--110},
    month = feb,
    year = {2014},
    issn = {1095-9572},
    doi = {10.1016/j.neuroimage.2013.10.067}
}

@article{Nagarajan2023Relevance-basedInterface,
    title = {{Relevance-based channel selection in motor imagery brain-computer interface}},
    year = {2023},
    journal = {Journal of Neural Engineering},
    author = {Nagarajan, Aarthy and Robinson, Neethu and Guan, Cuntai},
    number = {1},
    volume = {20},
    doi = {10.1088/1741-2552/acae07},
    issn = {17412552}
}

@article{Nagarajan2021InvestigationInterfaces,
    title = {{Investigation on Robustness of EEG-based Brain-Computer Interfaces}},
    year = {2021},
    journal = {Annual International Conference of the IEEE Engineering in Medicine and Biology Society. IEEE Engineering in Medicine and Biology Society. Annual International Conference},
    author = {Nagarajan, Aarthy and Robinson, Neethu and Guan, Cuntai},
    volume = {2021},
    doi = {10.1109/EMBC46164.2021.9630031},
    issn = {26940604}
}
